\documentclass[a4paper,twoside,twocolumn,english,amssymb,aps,prb,reprint,final,twocolumn,superscriptaddress,floats,showkeys,showpacs,citeautoscript]{revtex4-2}
\usepackage{graphicx,amsmath}
\usepackage{mathtools}
\usepackage{esvect}
\usepackage{epstopdf}
\usepackage{color}
\usepackage{float}
\usepackage[T1]{fontenc} 
\usepackage[utf8]{inputenc}
\usepackage{physics}
\usepackage{dcolumn}
\usepackage{bm}
\usepackage{accents}
\usepackage{gensymb}
\usepackage{siunitx}
\newcommand{\dbtilde}[1]{\accentset{\approx}{#1}}

\usepackage[]{newtxtext} 

\usepackage[dvipsnames]{xcolor}
\usepackage{hyperref}
\hypersetup{
	colorlinks,
	linkcolor={blue!90!black},
	citecolor={blue!90!black},
	urlcolor=	{blue!90!black}
}

\usepackage{placeins}

\begin{document}

\title{Dark-bright excitons mixing in alloyed InGaAs self-assembled quantum dots}
\author{Michał Zieliński}
\email{mzielin@fizyka.umk.pl}
\affiliation{Institute of Physics, Faculty of Physics, Astronomy and
Informatics\\
Nicolaus Copernicus University, ul. Grudziądzka 5, 87-100 Toruń, Poland}

\date{\today}

\begin{abstract}
Quantum dots are arguably one of the best  platforms for optically accessible spin based qubits. The paramount demand of extended qubit storage time can be met by using quantum-dot-confined dark exciton: a long-lived electron-hole pair with parallel spins. Despite its name the dark exciton reveals weak luminescence that can be directly measured. 
The origins of this optical activity remain largely unexplored.
In this work, using the atomistic tight-binding method combined with configuration-interaction approach, we demonstrate that atomic-scale randomness strongly affects oscillator strength of dark excitons confined in self-assembled cylindrical InGaAs quantum dots with no need for faceting or shape-elongation.
We show that this process is mediated by two mechanisms: mixing dark and bright configurations by exchange interaction, and equally important appearance of non-vanishing optical transition matrix elements that otherwise correspond to nominally forbidden transitions in a non-alloyed case.
The alloy randomness has essential impact on both bright and dark exciton states, including their energy, emission intensity, and polarization angle. 
We conclude that, due to the atomic-scale alloy randomness,
finding dots with desired dark exciton properties may require exploration of a large ensemble, similarly to how dots with low bright exciton splitting are selected for entanglement generation.
\end{abstract}

\maketitle

\section{Introduction}
The optically weakly active~\cite{Poem.Nature.2010,PhysRevX.5.011009,PhysRevB.95.165304,PhysRevLett.120.267401,PhysRevLett.107.166604} exciton states confined in quantum dots~\cite{arek-book,bimberg-book}, 
also known as dark excitons (DEs),~\cite{bayer-eh} are promising candidates for applications in quantum information processing.~\cite{PhysRevB.92.201201,PhysRevX.5.011009,Zielinski.PRB.2015,mcfarlane2009gigahertz,xu2008coherent,ohta2018dynamic}
Together with optically active bright excitons (BEs), electron-hole pairs with opposite spins far better studied for, e.g., entanglement generation~\cite{michler2017quantum,PhysRevLett.84.2513,Stevenson2006,prilmuller2018hyperentanglement}, they may form important building blocks for 
future quantum devices.
However, this applicability strongly depends on ability to understand and control the details of excitonic spectra, so-called ``excitonic fine structure''.~\cite{bayer-eh} 
On the one hand, the bright exciton spectrum is strongly determined by quantum-dot symmetry properties~\cite{bester-zunger-nair,karlsson,singh-bester-eh,Dupertuis.PRL.2011,zielinski-alloynwd,zielinski-elong,zielinski-elong2,swiderski-zielinski,ZielinskiSubstrate2012,GawelczykPRB2017,GawelczykAPPA2018,dash-mrowinski} with extensive efforts~\cite{abbarchi2010fine,kors2018telecom,young2005inversion,langbein2004control,bjork,borgstrom-nwqd,dalacu-selective,DalacuUltraClean,maaike-strain,yanase2017single,PhysRevB.84.235412,Swiderski.PhysRevB.100.235417,gong.PhysRevLett.106.227401,trotta2014highly} aimed at reduction of the bright exciton splitting (BES). 
On the other, the dark exciton fine structure has been studied to a far lesser degree. In particular, contrary to simplified theoretical predictions~\cite{bayer-eh}, the dark exciton can gain a non-negligible optical activity, even without an external magnetic field,~\cite{karlsson,zielinski-alloynwd} whereas origins of this luminescence remain largely unexplored.~\cite{don2016optical,germanis.PhysRevB.98.155303}

Accurate theoretical modeling of excitonic fine structure, regarding both bright and dark excitons, still presents a serious challenge for approaches utilizing continuum-media approximation,~\cite{kp14,kp14more,karlsson} and even for atomistic methods.~\cite{singh-bester-ordering,zielinski-alloynwd,Zielinski.PRB.2015}
In this work, we use the combination of tight-binding (TB) and configuration-interaction (CI) methods, which proved its ability to find excitonic fine structure in a good agreement with experiment.~\cite{bryant2013mechanism,zielinski-vbo,zielinski-alloynwd}
To understand the role of disorder due to alloy randomness, we present an extensive theoretical study of properties of bright and dark excitons calculated for an ensemble of 300 alloyed In$_{0.5}$Ga$_{0.5}$As self-assembled quantum dots, each treated with the same atomistic resolution.

We find that alloy randomness, by reducing the overall symmetry, leads to (i) non-vanishing matrix elements of the optical transition (optical/transition dipole moments), as well as (ii) non-negligible exchange integrals mixing bright and dark excitonic configurations.
There are thus two equally important sources of non-zero oscillator strength for the in-plane dark excitonic emission. 
On the contrary, the out-of-plane ($z$) polarized emission is dominated by the contribution from the matrix element of the optical transition only, which is in turn governed by the valence-band mixing, in agreement with the phenomenological understating of dark exciton states. Therefore, contrary to the case of in-plane emission, the exchange mixing does not play a role for $z$-polarized dark exciton emission.

While most of the quantum dots in ensembles reveal rather weak dark exciton activity, we show that a mere alloy randomness triggers the dark exciton optical activity reaching up to 1/6000 fraction of that for the bright exciton, without faceting or shape elongation.~\cite{PhysRevB.95.165304,Zielinski.PRB.2015,germanis.PhysRevB.98.155303} This conclusion is valid for both out-of-plane and in-plane polarizations. As optical activity of dark exciton is inversely proportional to its lifetime, a strong variation of the former translates to a pronounced dot-to-dot variations of the latter on a millisecond scale, while the bright exciton lifetimes tend to be systematically close to 1~ns. Moreover, despite the overall cylindrical quantum-dot shape, and despite strong alloying, the polarization properties of both bright and dark excitons for in-plane emission reveal hallmarks of a pronounced anisotropy due to the underlying crystal lattice.

The paper is organized as follows: after a short theoretical introduction in Section~\ref{section:methods},
we start with the discussion of bright exciton spectra and polarization angle in Section~\ref{section:results},
Throughout most of the paper we aim to bridge between atomistic results, and the effective treatment based on valence-band mixing. When feasible these discussions are augmented with an abridged statistical analysis.
In Section~\ref{section:dark} we study in-plane emission form dark excitons, including their polarization properties, and finally we study excitonic lifetimes in Section~\ref{section:lifetime}.

\section{Systems and methods}
\label{section:methods}

The calculation starts with finding atomic positions that minimize the total elastic energy, by using the valence force field (VFF) method of Keating~\cite{keating,martin}. Minimization of strain energy is performed with the conjugate gradient method.~\cite{jaskolski-zielinski-prb06,saito-arakawa}
Next, the piezoelectric potential~\cite{PhysRevLett.96.187602,PhysRevB.74.081305,PhysRevB.84.195207,tse2013non,Caro.PhysRevB.91.075203,swiderski2016exact} is calculated by accounting for both linear and quadratic contributions, with piezoelectric coefficients from Ref.~\cite{PhysRevB.84.195207}. The single-particle spectra of electrons and holes are obtained with the empirical $sp^3d^5s^*$ tight-binding method accounting for $d$-orbitals and spin-orbit interaction.~\cite{jancu,chadi-so-in-tb,zielinski-including,zielinski-vbo}
The tight-binding calculation is performed on a smaller domain than the valence force field calculation.~\cite{lee-boundary,zielinski-multiscale}
The details regarding the $sp^3d^5s^\star$ tight-binding calculations for various nanostructures are discussed thoroughly in our earlier papers.~\cite{zielinski-including,zielinski-vbo,jaskolski-zielinski-prb06,zielinski-prb09,PhysRevB.75.245433} 

We note that eigenstates of the tight-binding Hamiltonian are doubly degenerate to due to time reversal-symmetry.~\cite{dresselhaus2007group} 
Therefore, for convenience, in several places we label tight binding electron ground states with up and down spin-indices, i.e., $e_{\uparrow}$ and $e_{\downarrow}$. 
However, we emphasize that due to spin-orbit interaction, and low overall symmetry, due to underlying lattice with alloying, the spin in no longer a good quantum number, and such labeling is only approximate.
Similarly, we label atomistically obtained hole ground states with double-arrows, i.e., $h_{\Uparrow}$ and $h_{\Downarrow}$, even though due to valence-band mixing and alloying they do not exactly correspond to heavy-hole $\pm\frac{3}{2}$ eigenstates.

After the tight-binding stage of calculations, the excitonic spectra~\cite{michler} are calculated with the configuration-interaction method described in detail in Ref.~\cite{zielinski-prb09}. 
The Hamiltonian for the interacting electron-hole pair can be written in second quantization as follows~\cite{michler}
{\begin{multline}
  \hat{H}_{ex} =  \sum_{i}E_i^ec_i^\dagger c_i+\sum_{i}E_i^hh_i^\dagger h_i  \\
  -\sum_{ijkl}V_{ijkl}^{eh,\text{dir}} c_i^\dagger h_j^\dagger h_k c_l
  +\sum_{ijkl}V_{ijkl}^{eh,\text{exch}} c_i^\dagger h_j^\dagger c_k h_l
  \label{CIHamiltonian}
\end{multline}}
where $E_i^e$ and $E_i^h$ are the single-particle electron and hole energies
obtained at the single-particle stage of calculations, respectively, and
$V_{ijkl}$ are Coulomb matrix elements (Coulomb direct and exchange integrals) calculated according to the procedure given in Ref.~\cite{zielinski-prb09}. 
More details regarding the Coulomb matrix element computation for tight-binding wave-functions can also be found in Refs.~\cite{lee-johnsson-prb2001,schulz-schum-czycholl} as well as in our recent papers.~\cite{rozanski-zielinski,rozanski2019efficient}

Finally, the optical spectra are found by calculating the
intensity of photoluminescence from the recombination of
electron-hole pair using Fermi's golden rule \begin{equation}
\begin{split}
I\left(\omega\right)=\Big|\sum_{c,v}C_{c,v}^i
\matrixel{\Psi_{c}}{\vec{\boldsymbol{\epsilon}}\cdot\vec{\boldsymbol{r}}}{\Psi_{v}}\Big|^2
\delta\left(E_i-\si{\hbar}\omega\right)\\
=\Big|\sum_{c,v}C_{c,v}^i M_{c,v}^{\vec{\boldsymbol{\epsilon}}}\Big|^2
\delta\left(E_i-\si{\hbar}\omega\right)
\end{split}
\label{Eq:intensity}
\end{equation}
where $E_i$ is the energy of initial $i$th state of the exciton, $C_{v,c}^i$ are CI expansion coefficients for the $i$th state obtained by solving Eq.~\ref{CIHamiltonian},
$M_{c,v}^{\vec{\boldsymbol{\epsilon}}}\equiv\matrixel{\Psi_{c}}{\vec{\boldsymbol{\epsilon}}\cdot\vec{\boldsymbol{r}}}{\Psi_{v}}$ is the optical dipole moment matrix element calculated from
single-particle tight-binding electron $\Psi_{c}$ and hole $\Psi_{h}$ wave functions respectively, for a given polarization of light $\vec{\boldsymbol{\epsilon}}$, e.g., 
$M_{e_{\uparrow},h_{\Downarrow}}^{z}=\matrixel{e_{\uparrow}}{z}{h_{\Downarrow}}$, for $z$-polarized (out-of-plane) light, and optical (transition) dipole matrix element involving ground electron and hole states with opposite (quasi-)spins.

Fermi’s golden rule allows for calculation of first-order radiative recombination lifetime in a similar manner~\cite{karrai2003optical,PhysRevB.72.245318,oliveira}
\begin{equation}
\frac{1}{\tau_i}=\frac{4n\alpha\omega_i^3}{3c^2}
\sum_{\vec{\boldsymbol{\epsilon}}=
\vec{\boldsymbol{x}},
\vec{\boldsymbol{y}},
\vec{\boldsymbol{z}}
}
\Big|\sum_{c,v}C_{c,v}^i M_{c,v}^{\vec{\boldsymbol{\epsilon}}}\Big|^2,
\label{Eq:lifetime}
\end{equation}
where $\omega_i=E_i/\hbar$, $n=3.6$~\cite{karrai2003optical} is the refractive index, $\alpha$ is the fine-structure constant, $c$ is the speed of light in the vacuum, $x$, $y$ are in-plane, and $z$ is the out-of-plane radiative decay channel.
Finally, the average bright exciton radiative lifetime, assuming identical occupation of two bright exciton states BE1 and BE2 (discussed later in the text), and neglecting thermal effects is calculated as~\cite{PhysRevB.72.245318,oliveira}
\begin{equation}
\frac{1}{\tau_{BE}}\approx\frac12 \left(\frac{1}{\tau_{BE1}}+\frac{1}{\tau_{BE2}}\right)
\label{Eq:lifetime2}
\end{equation}
with an analogous formula for the dark exciton.

\begin{figure}
 \begin{center}
  \includegraphics[width=0.48\textwidth]{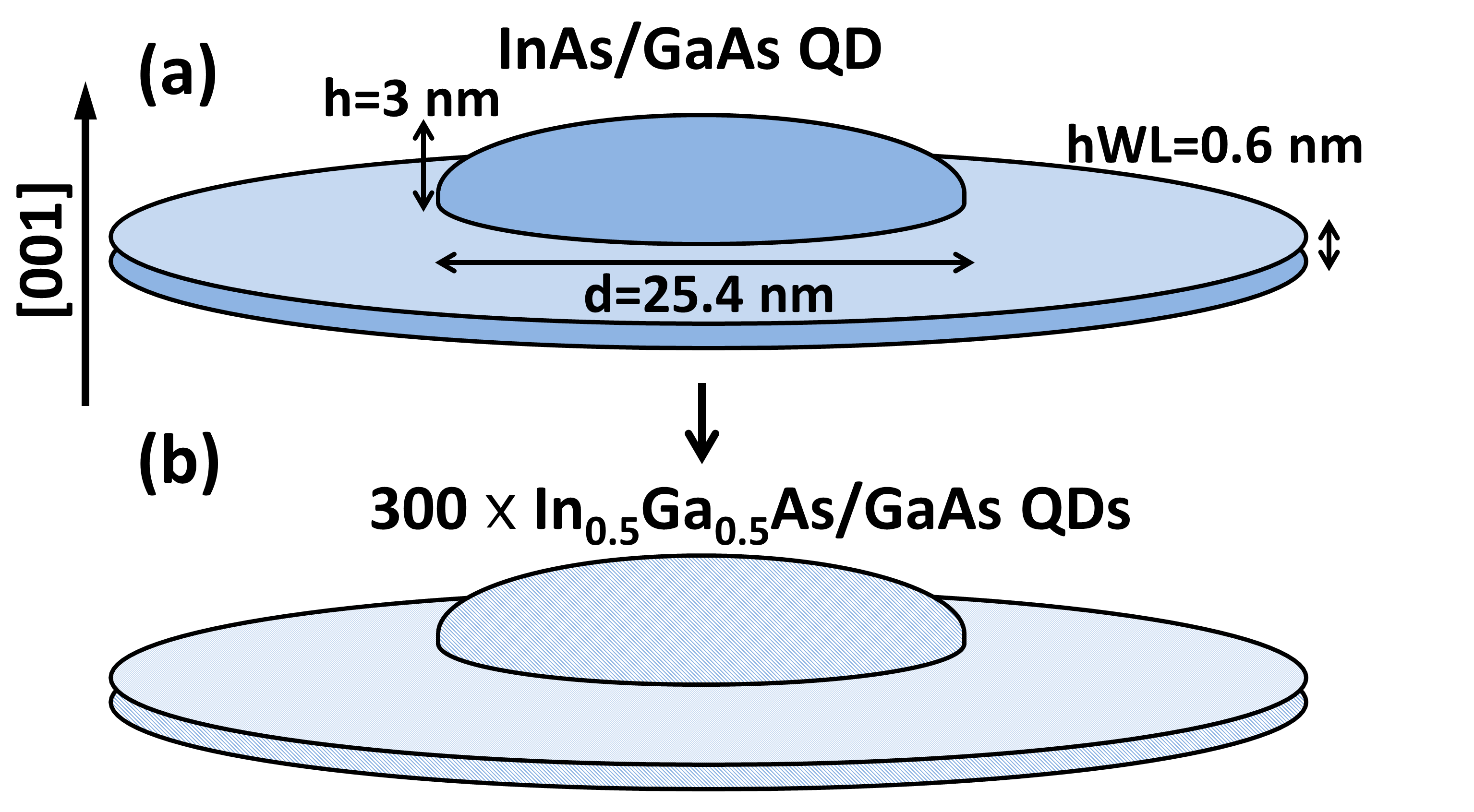}
  \end{center}
  \caption{Schematics of systems under consideration: (a) lens-shaped InAs/GaAs self-assembled quantum dot located on a wetting layer and (b) alloyed In$_{0.5}$Ga$_{0.5}$As quantum dots with the same dimensions, however with 5 randomly generated samples (realizations) corresponding to the same average composition, yet different (random) atomic arrangement. See the text for details. Surrounding GaAs material is not shown.}
  \label{schematics}
\end{figure}

Using the above formalism, we perform calculations for an ensemble of alloyed In$_{0.5}$Ga$_{0.5}$As lens-shaped quantum dots located on a wetting layer (Fig.~\ref{schematics}), with low overall (shape + alloyed lattice) $C_{1}$ symmetry.
Each quantum-dot system is treated separately by full VFF/TB/CI calculation.
The height of all quantum dots is equal to 3~nm, whereas 
the diameter is 25.4~nm. Each quantum dot is located on a 0.6~nm (1 lattice constant) thick wetting layer.
A quantum-dot system with such dimensions (in particular in a non-alloyed InAs/GaAs variant), has been studied thoroughly in literature (including our own work~\cite{zielinski-including,zielinski-vbo}), as a model of a typical self-assembled quantum dot.
For comparison, in several places we discuss as well results obtained for $C_{2v}$ InAs quantum, with no gallium addmixture
(no alloying; upper part of Fig.~\ref{schematics}).
However, our focus is on the alloyed system, by considering 50\% addmixture of barrier material in the dot region, i.e., an In$_{0.5}$Ga$_{0.5}$As quantum dot in the GaAs surrounding, with uniform (on average) composition profile. We consider 300 random samples (lower part of Fig.~\ref{schematics}), i.e., 300 different random realizations of the same alloyed lens-shaped quantum dot, with nominally the same average composition yet different atomic arrangement on a microscopic scale. 

\section{Results}
\label{section:results}
\begin{figure}
 \begin{center}
  \includegraphics[width=0.48\textwidth]{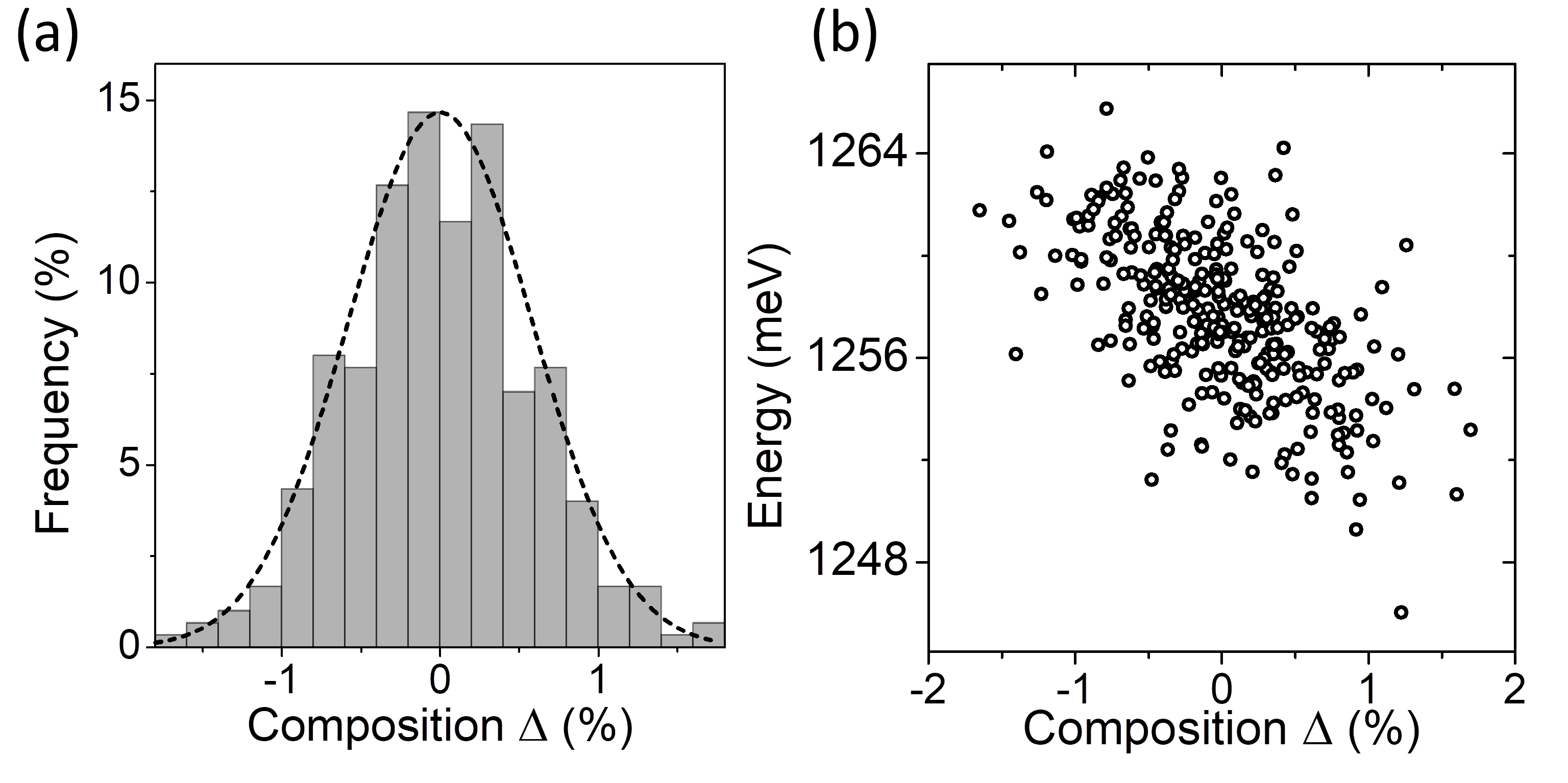}
  \end{center}
  \caption{(a) Histogram illustrating distribution of average composition variations (from ideal 50\% gallium content) for a family of 300 alloyed quantum dots studied in the text; (b) exciton ground-state energy as function of composition difference and alloy randomness. See the text for details.}
  \label{fig:Xenergy}
\end{figure}

To study effects of alloy randomness, we focus on uniform composition profile, rather than on effects of spatial changes in the overall composition.~\cite{PhysRevB.79.125316,PhysRevLett.84.334,Zunger-EPM2}
To model random alloy fluctuations, for each cation site a random number (uniformly distributed within 0 to 1 range) is generated, and based on that Ga atom is replaced with In with 0.5 probability. This approach will thus generate microscopic configurations that differ significantly in the atomic arrangement, while maintaining (on average) 50\% gallium content. A more careful inspection reveals however, that such procedure may in fact result in random realizations that differ from average composition within approximately $\pm1$\% range, as shown in Fig.~\ref{fig:Xenergy} (a). Therefore, a generated ensemble of quantum dots is a subject to (a) randomness due to different atomic arrangement corresponding to the same composition, plus (b) small ($\pm\sim$1\%) fluctuations of average composition.
As shown in Fig.~\ref{fig:Xenergy} (b) the ground state excitonic energy of the considered ensemble of quantum dots is altered by both effects.
Changes in average composition ($x$ axis in Fig.~\ref{fig:Xenergy}) correspond to a change of excitonic energy by approximately 8 meV per 1\% of gallium content, consistent with 10 meV estimate that can be obtained from virtual crystal approximation (VCA) with bulk InGaAs parameters taken from Ref.~\cite{vurgaftman2001band}
However, changes on the vertical axis, which are also on a scale of several meV's, are related purely due to alloy randomness, that is going beyond the VCA.
Importantly, whatever the cause, the exciton ground energy varies within relatively small ranges, namely less then 1.3\% from the average of approximately 1257.6~meV.
Therefore, as a rule of thumb, any spectral quantity that changes substantially more than 2\% when going from a sample to sample (such as the BES and the DES discussed in the following) cannot be attributed to a change in averaged composition, but to alloy randomness (or changes in the microscopic atomic arrangement).

\begin{figure*}
 \begin{center}
  \includegraphics[width=0.80\textwidth]{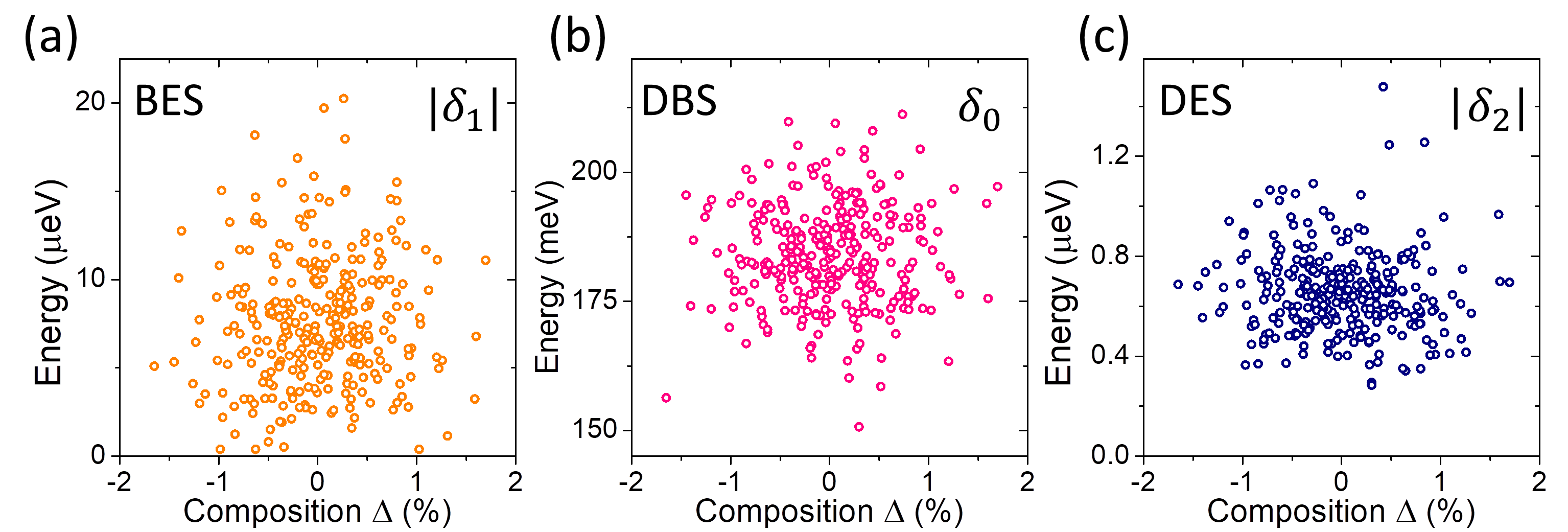}
  \end{center}
  \caption{Bright exciton splitting (a), dark-bright splitting (b), and dark exciton splitting (c) as a function of difference from average composition ($x$ axis), and alloy randomness (distribution of values along $y$ axis). See the text for more details.}
  \label{fig:splittings}
\end{figure*}

In a standard treatment, the excitonic fine structure is typically~\cite{bayer-eh,Ivchenko.Book.2005} addressed by starting from idealized single-particle states, with electron spin states $\ket*{\pm\frac{1}{2}}$, and the hole ground states of heavy-hole character,
associated with the projection of the total hole angular
momentum: $\ket*{\pm\frac{3}{2}}$.
It is important to reemphasize that $\ket*{\pm\frac{1}{2}}$ are not exactly equal to $e_{\uparrow}/e_{\downarrow}$ tight-binding eigenstates, which are not eigenfuctions of spin, but are labeled with pseudo-spin index instead.
The same applies the $h_{\Uparrow}/h_{\Downarrow}$ hole states as well. Nevertheless, $\ket*{\pm\frac{1}{2}}$, $\ket*{\pm\frac{3}{2}}$ states are of vital importance for defining effective Hamiltonians.~\cite{bayer-eh,tsitsishvili2017impact} 
To this end the excitonic basis of four states in such approaches is thus constructed from the above-mentioned, idealized single-particle states, and characterized by their angular momentum projections
\begin{equation}
\left|\pm 1\right>=\left|\pm\frac{3}{2},\mp\frac{1}{2}\right>,\left|\pm 2\right>=\left|\pm\frac{3}{2},\pm\frac{1}{2}\right>,
\label{basis}
\end{equation}
where the first value in the ket corresponds to the hole, and the second one to the electron.
Matrix elements of the optical transition are given as:~\cite{Ivchenko.Book.2005}
$M_{\pm\frac{3}{2},\mp\frac{1}{2}}\equiv M_{\pm 1} =M\left(e_x \mp ie_y\right), M_{\pm 2} = 0$, where $e_x, e_y$ are in-plane components of $\mathbf{e}$ polarization vector, and $M$ is independent of the light polarization. 
Therefore $\left|\pm 1\right>$ are optically active, whereas $\left|\pm 2\right>$ are optically inactive states.

Electron-hole exchange Hamiltonian is expanded in basis of these states, and in the order
$\left|1\right>,\left|-1\right>,\left|2\right>,\left|-2\right>$ can be written as~\cite{bayer-eh}
\begin{align}
\mathbf{H}_{\mathrm{exch}} &=\frac{1}{2}
\begin{bmatrix}
    \delta_{\mathrm{0}} & \delta_1  & 0 & 0  \\
    \delta_1  & \delta_{\mathrm{0}} & 0 & 0  \\
    0 & 0 & -\delta_{\mathrm{0}} & \delta_2  \\
    0 & 0 & \delta_2  & -\delta_{\mathrm{0}}
\end{bmatrix},
\label{Hexch}
\end{align}
where $\delta_0$ describes the dark-bright exciton splitting (``isotropic electron-hole exchange interaction''), $\delta_1$ is responsible for the bright doublet splitting, and since it is related to distortions from idealized symmetry, it is often called ``anisotropic electron-hole exchange interaction'', and finally $\delta_2$ is responsible for the dark exciton splitting. For $C_{3v}$, and $D_{2d}$ high-symmetry quantum dots
$\delta_1=0$, i.e., bright exciton splitting vanishes.~\cite{singh-bester-eh} For $C_{3v}$ systems additionally $\delta_2=0$.
Nonetheless, for realistic alloyed $C_1$ quantum dots both $\delta_1\neq0$ and $\delta_2\neq 0$, and $\delta_1 \gg \delta_2$. Similar properties are shared by $C_{2v}$ quantum dots, hence $C_{2v}$ systems are often treated as good approximations to real system, although such approximation often fails,~\cite{singh-bester-lower,zielinski-elong3} as also discussed in this work.

Since the off-diagonal block consists of zero's, there is no coupling between bright and dark manifolds, and they can be treated separately, with bright exciton states given as symmetric/anti-symmetric combinations of basis states
\begin{align}
\ket*{BE1}&=\left(\ket*{1}-\ket*{-1}\right)/\sqrt{2}\\
\ket*{BE2}&=\left(\ket*{1}+\ket*{-1}\right)/\sqrt{2},
\end{align}
with analogous formulas for the dark exciton.
Corresponding eigenenergies are given as
\begin{align}
E_{BE1}&=\left(\delta_0-\delta_1\right)/2\\
E_{BE2}&=\left(\delta_0+\delta_1\right)/2,
\end{align}

For sake of comparison, atomistic results (calculated in a basis of 144 CI configurations) can be ``recast'' to spectra of Eq.\ref{Hexch}. 
For non-alloyed, lens-shapes system of $C_{2v}$ symmetry we found from atomistic simulations, that $\delta_1>0$, thus in such a case anti-symmetric $\ket*{BE1}$ would be the lower-energy bright excitonic state. Moreover, $\ket*{BE1}$ is linearly polarized along $[1\overline{1}0$ direction, and $\ket*{BE2}$ is linearly polarized along $[110]$. Notably, there is no $z$-polarized emission allowed. For an alloyed system polarization properties are more complicated, and will be discussed in the following part of the paper.

The bright and dark exciton splittings calculated atomistically correspond to $\left|\delta_1\right|$, $\left|\delta_2\right|$ of the simple model respectively, which are shown in Fig.~\ref{fig:splittings}.
Excitonic fine structure does not show any clear trend with respect to average composition fluctuations, yet, it strongly varies due to alloy randomness.~\cite{gong.PhysRevB.89.205312,zielinski-alloynwd}
Bright exciton splitting [Fig.~\ref{fig:splittings} (a)] varies from zero to over 20~$\mu$eV, with average value of 7.6~$\mu$eV, and standard deviation of 3.6~$\mu$eV.
The dark-bright splitting [Fig.~\ref{fig:splittings} (b)] has the average value of 184~$\mu$eV, whereas the dark exciton splitting varies
from 0.28 to 1.48~$\mu$eV, with mean value of 0.66~$\mu$eV and standard deviation equal to 0.16~$\mu$~eV. 
Again, since we do considered cylindrical quantum dot shape shape~\cite{zielinski-elong3} both BES and DES are in good agreement with experimental values, although DES results are generally smaller than observed in the experiment.~\cite{PhysRevB.92.201201}
Nevertheless, we note that this range of DE energy level splittings corresponds to the precession periods~\cite{Poem.Nature.2010} of a coherent superposition of DE eigenstates varying from
to 2.8 to 14.5~ns, (with average of 6.3~ns), thus again overlapping with typical experimental values.~\cite{PhysRevB.92.201201,heindel2017accessing}

To further analyze polarization properties of bright excitons, while still assuming no dark-bright exciton coupling, one can use the Hamiltonian for the bright manifold only, derived based on group-theoretical~\cite{gong.PhysRevLett.106.227401} arguments:
\begin{align}
\mathbf{H} &=
\begin{bmatrix}
    E+\delta & \kappa  \\
    \kappa  & E-\delta
\end{bmatrix},
\label{HBESatom}
\end{align}
where $E$ is the reference excitonic energy, $\delta$ (with no subscript) is responsible for the bright exciton splitting in the $C_{2v}$ case due to lattice (and shape) anisotropy, whereas $\kappa$ determines the fraction of the BES related to contribution due to lowering of symmetry caused by alloying. Moreover, $\kappa$ is responsible for the coupling between the two bright states, leading to the rotation~\cite{gong.PhysRevLett.106.227401}
of the emission lines from the perfect crystal directions. This polarization angle $\theta$ is given as:
$\mathrm{tan}\left(\theta\right)=\frac{\delta}{\kappa}\pm\sqrt{1+\left(\delta/\kappa\right)^2}$, with the overall bright exciton splitting given as $\delta_{\mathrm{1}} =2 \sqrt{\delta^2+\kappa^2}$, and conversely
\begin{equation}
\kappa = -\frac{\delta_1}{2}\mathrm{sin}\left(2\theta\right),\;
\delta = \frac{\delta_1}{2}\mathrm{cos}\left(2\theta\right)
\end{equation}
with bright eigenstates given as 
\begin{align}
\ket*{BE_{1}}&=-\ket*{\mathrm{sin}\left(\theta\right)}+ \ket*{\mathrm{cos}\left(\theta\right)}\\
\ket*{BE_{2}}&=\ket*{\mathrm{cos}\left(\theta\right)}+ \ket*{\mathrm{sin}\left(\theta\right)}
\end{align}

\begin{figure}
 \begin{center}
  \includegraphics[width=0.48\textwidth]{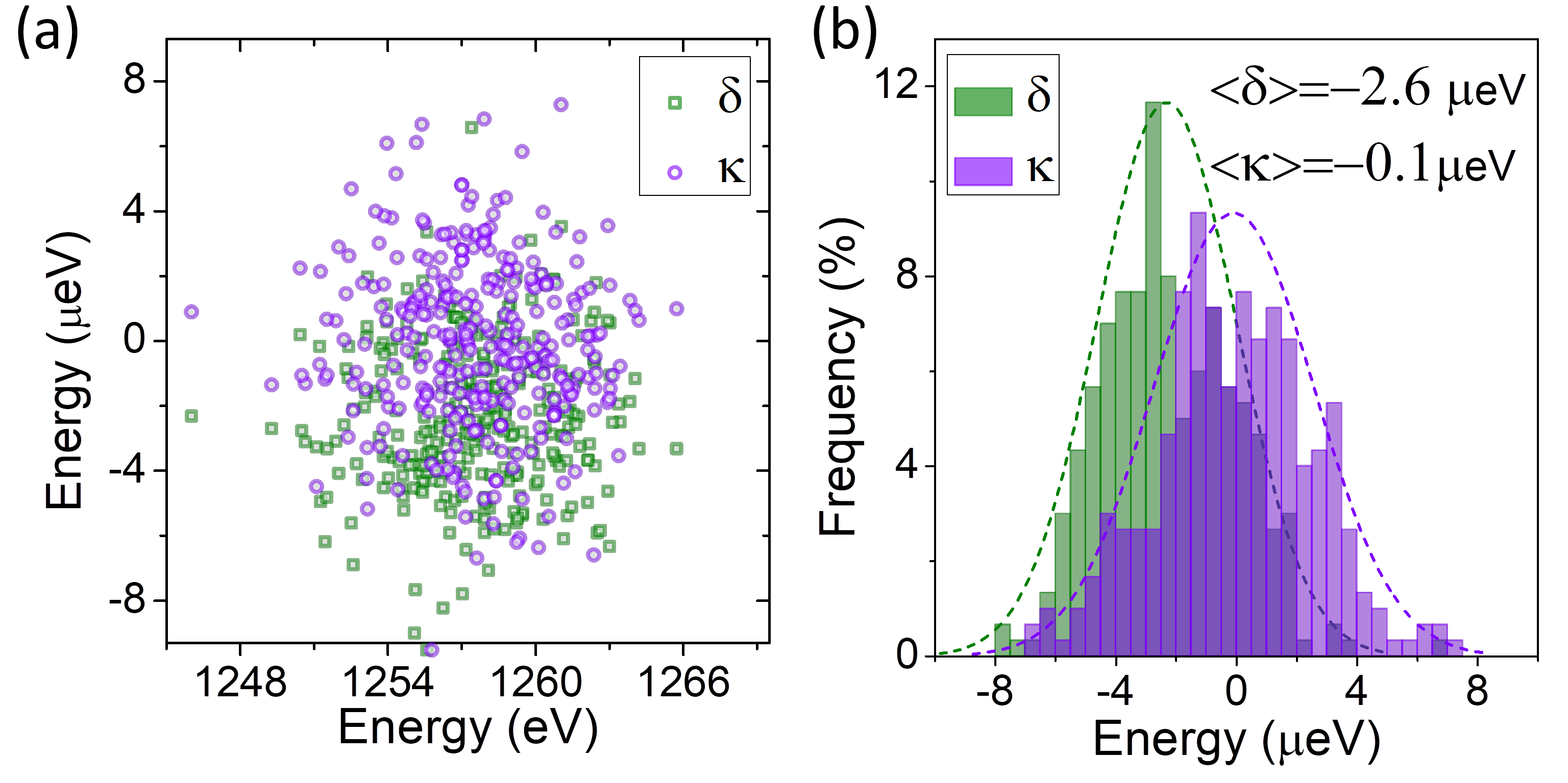}
  \end{center}
  \caption{(a) $\delta$ and $\kappa$ contribution to the BES as function of excitonic ground state energy. Histogram (b) of $\delta$ and $\kappa$ with mean $\kappa$ (contribution due to alloyin) value close to zero, mean  $\delta$ (contribution due to lattice anisotropy) different from zero. See the text for details.}
  \label{fig:delta_kappa}
\end{figure}

Model Eq.\ref{HBESatom} allows to decompose the bright exciton splitting into two contributions, due to anisotropy ($\delta$) and alloying ($\kappa$) respectively, which are shown in Fig.~\ref{fig:delta_kappa}.
The $\kappa$ mean value, $\left<\kappa\right>$, is close to zero indicating lack of any directional character of alloying, whereas $\delta$ has mean value of -2.32~$\mu$eV, which is due lattice anisotropy. This result is in vivid contrast to atomistic pseudopotential calculations,~\cite{gong.PhysRevB.89.205312} where $\left<\delta\right>$ is different from zero only for elongated quantum dots. This result is however consistent with systematic difference between empirical tight-binding and empirical pseudopotential approaches for $C_{2v}$ systems,~\cite{zielinski-vbo}
as well BES reported experimentally.~\cite{singh-bester-ordering}

\begin{figure}
 \begin{center}
      \includegraphics[width=0.48\textwidth]{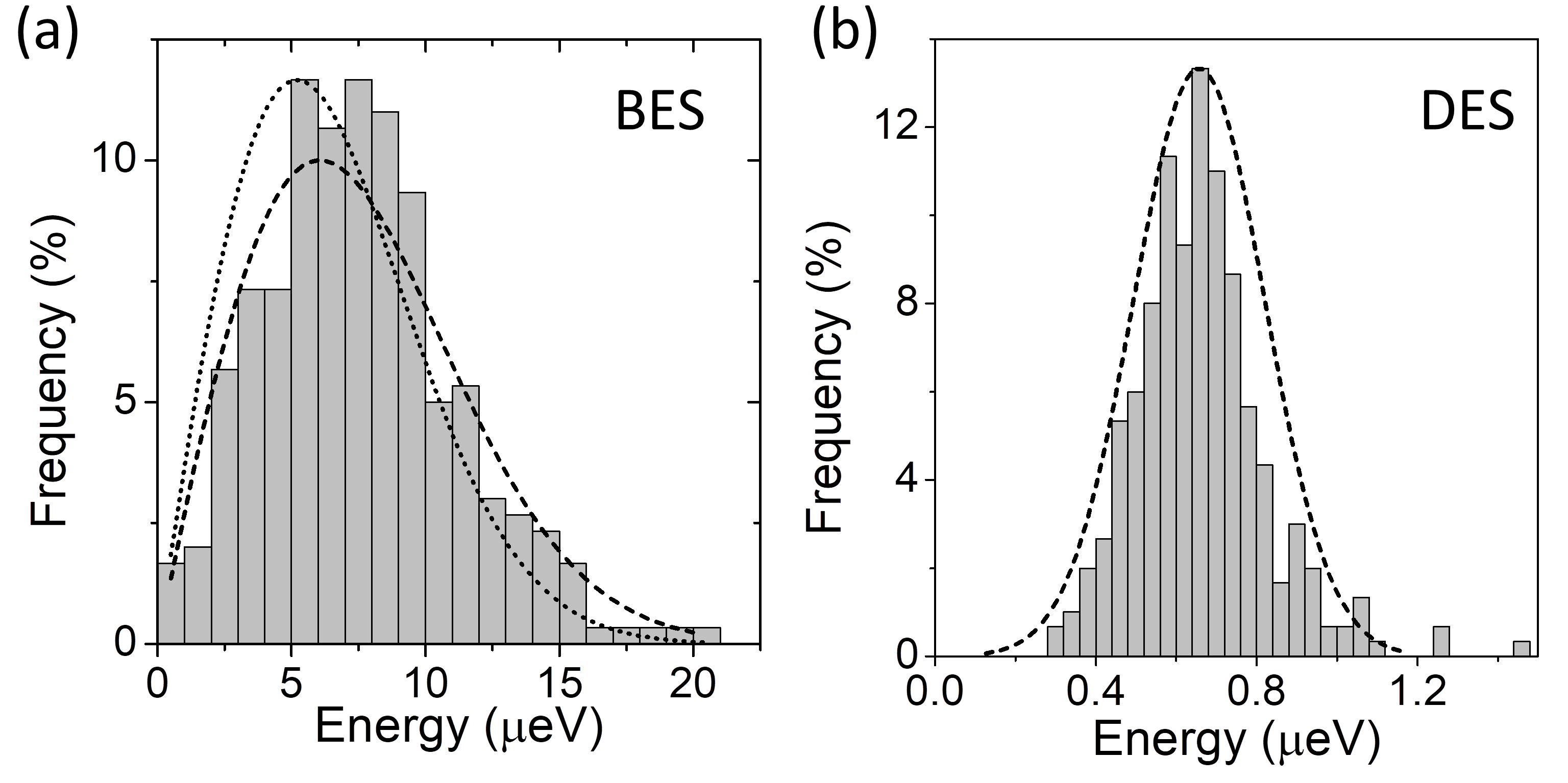}
  \end{center}
  \caption{Histogram of (a) the bright exciton splitting (with dotted/dashed line corresponding to Rayleigh/Wigner distributions respectively), and (b) dark exciton splitting (the dashed line is a fit to a normal distribution).}
  \label{fig:besdeshist}
\end{figure}

$\kappa$ and $\delta$ have standard deviations $\sigma$ of $2.66~\mu$eV, and 2.30~$\mu$eV, respectively.
Since both $\kappa$ and $\delta$ are apparently independent, and have distribution that can be quite well described by the normal distribution [Fig.~\ref{fig:delta_kappa} (a)], with
 $\sigma=\sigma_{\delta}\approx \sigma_{\kappa}=2.6~\mu$eV, and $\left<\delta\right>$ not exceeding $\sigma$, one can expect that 
 $\delta_{\mathrm{1}} =2 \sqrt{\delta^2+\kappa^2}$, will follow Rayleigh distribution,~\cite{wojnar2013rayleigh,mehta2004random} (i.e., essentially a $\chi$ distribution with two degrees of freedom):
\begin{equation}
P\left(\delta_1\right)=\frac{\delta_1}{4\sigma^2}\mathrm{exp}\left(-\frac{\delta_1^2}{8\sigma^2}\right),
    \label{eq:rayleigh}
\end{equation}
where the denominator is multiplied by a factor of 4 (with respect to the original Rayleigh distribution), due to 2 occurring in the definition of $\delta_{\mathrm{1}}$, i.e, $\delta_{\mathrm{1}} =2 \sqrt{\delta^2+\kappa^2}$.
Fig.~\ref{fig:besdeshist} (a) shows a histogram of BES values with Eq.~\ref{eq:rayleigh} plotted with dotted black line revealing a rather good fit, despite not strictly fulfilling the assumption of $\langle\delta\rangle=0$, and with $\sigma=2.66$, not taken as fitting parameter.
Thus, as a result of $\delta_{\mathrm{1}}$ having a two-component character its distribution varies from the normal distribution, but resembles $\chi$ with k=2.
Rayleigh distribution is derived for the system of two-dimensions, however the mathematical form of the Rayleigh distribution is identical with Wigner surmise for the one-dimension energy levels problem~\cite{mehta2004random,gong.PhysRevB.89.205312} 
\begin{equation}
P\left(s\right)=\frac{\pi}{2}s\,\mathrm{exp}\left(-\frac{\pi}{4}s^2\right),
\label{eq:wigner}
\end{equation}
with $s=\delta_1/\expval{\delta_1}$, and
where $\expval{\delta_1}=$ 7.60~$\mu$eV was calculated as an average BES of all 300 samples.
The plot of Eq.~\ref{eq:wigner} is also shown in Fig.~\ref{fig:besdeshist} (a) with a dashed line, compared with the histogram of BES values.
We note that Eq.\ref{eq:rayleigh} and Eq.~\ref{eq:wigner} have identical mathematical forms, whereas the difference between plots on Fig.~\ref{fig:besdeshist} (a) is related to difference in parameters describing ``variances'' of the both distributions (i.e., $\langle\delta_1\rangle\neq \sqrt{2\pi}\sigma$).

Since Eq.~\ref{eq:wigner} corresponds to the standard level-spacing distribution of the Gaussian orthogonal ensemble in random matrix theory, Gong et al.~\cite{gong.PhysRevB.89.205312} argue this sort of statistical dependence reveals strong repulsion between excitonic levels, and the presence of multiplicative $s$ in Eq.~\ref{eq:wigner} reduces the probability of finding quantum dots with small BES.
As a result of this sort of behavior, we find that over 50\% of samples has the BES within 5--10~$\mu$eV, with only in 3.4\% of cases with BES larger than 15~$\mu$eV, and importantly only 1.7\% cases with the BES smaller than 1~$\mu$eV.
BES smaller than the line-width (of approximately 1~$\mu$eV) is important in the context of entanglement generation.
Results presented here may look counter-intuitive, since cylindrical quantum dots seem to be good candidates for small BES values, yet are consistent with earlier findings.~\cite{gong.PhysRevB.89.205312}

For comparison Fig.~\ref{fig:besdeshist} (b) shows a histogram of the DES values revealing normal distribution with the mean of 0.657 and standard deviation of 0.16~$\mu$eV.
Thus, the DES is order of magnitude smaller than the BES, however, in systems studied here it is virtually impossible to find a quantum dot with a vanishing dark exciton splitting.

Finally, we note that mixing with higher shells increases both BES and DES. 
Should only the s-shell be included in the atomistic calculation (corresponding to $4\times4$ CI Hamiltonian) the mean BES value would be 4~$\mu$eV, and DES mean value 0.53~$\mu$eV, with no substantial qualitative difference in histograms. For comparison, atomistic results obtained for the s-shell only are presented in the Appendix A.

\begin{figure}
 \begin{center}
  \includegraphics[width=0.48\textwidth]{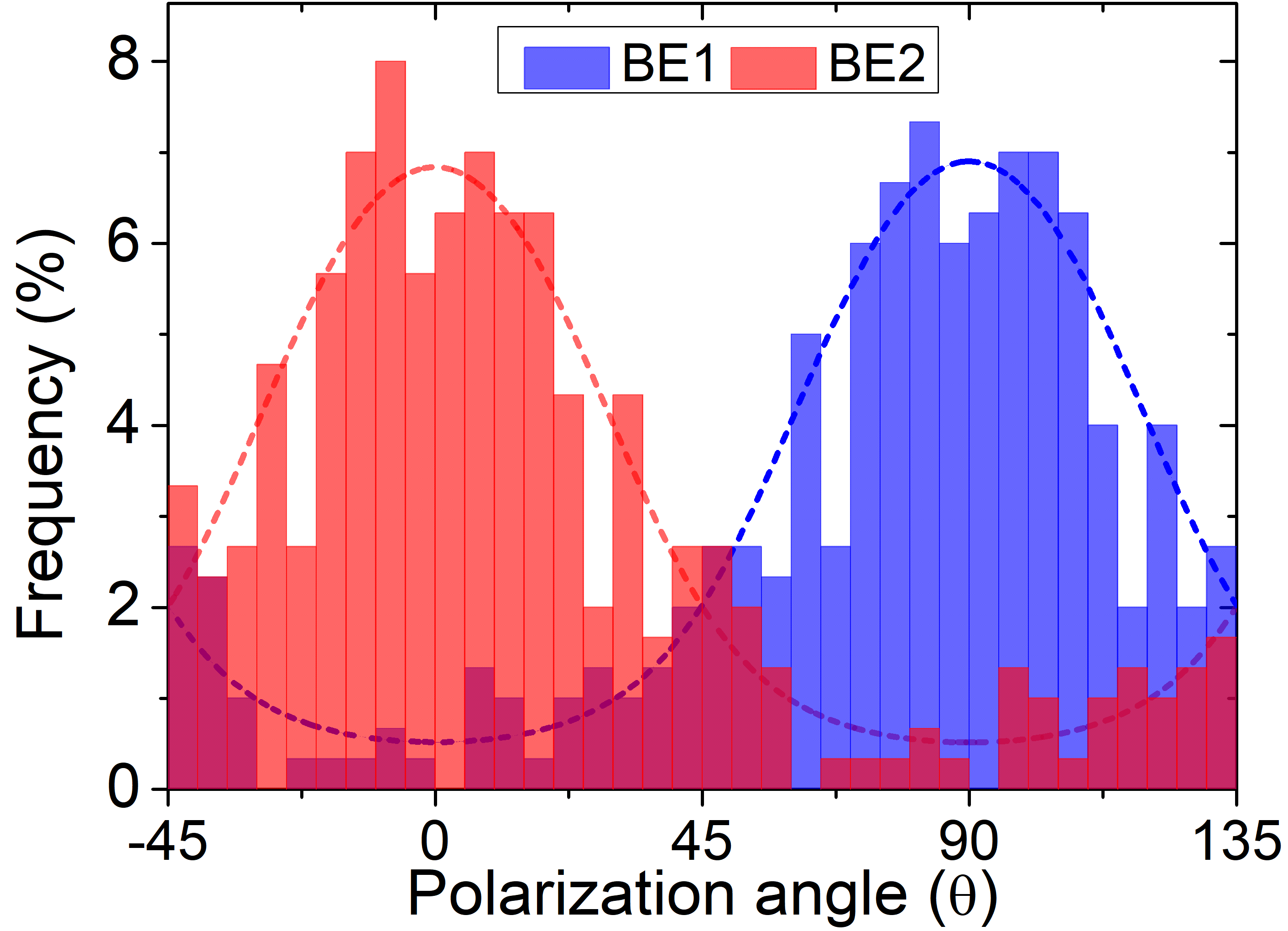}
  \end{center}
  \caption{Histograms of polarization angle for lower (BE1) and higher (BE2) energy bright exciton states with maxima corresponding to $[110]$ and $[1\underline{1}0]$ crystal axes (0 and 90 degrees respectively).
  The dashed lines are calculated using approach from Ref.~\cite{gong.PhysRevB.89.205312}
  See the text for details.}
  \label{fig:anglehist}
\end{figure}

Statistical analysis performed for $\kappa$ and $\delta$ can be performed for the bright exciton polarization angle as well, with results shown in Fig.~\ref{fig:anglehist}, where the histogram obtained from the current atomistic calculation is compared with a statistical model (dashed line) from Supplementary Information of Ref.~\cite{gong-prb77}.
The lower energy bright exciton state (BE1) tends be polarized along the [1$\overline{1}$0] direction (rotated by 90$\degree$ from [110]), whereas the higher-energy one (BE2) prefers polarization along the [110] direction. Therefore, despite substantial alloying (and overall cylindrical shape) polarization angle dependence has well-defined maxima corresponding to the vicinity of crystal axis.
This is again in stark contrast to empirical pseudopotential results~\cite{gong-prb77} that predict pronounced polarization angle distribution maxima only for elongated dots. This difference can be traced back to $\delta$ distribution as shown earlier (Fig.~\ref{fig:delta_kappa}): in atomistic tight-binding calculation $\langle\delta\rangle \neq 0$, contrary to the empirical pseudopotential method. 
This is consistent with the latter systematically underestimating lattice BES contribution, origin of which remains to a large degree unexplored.~\cite{singh-bester-ordering}

Model of Eq.~\ref{HBESatom} assumes the angle between bright exciton to be exactly equal to 90$\degree$.
To go beyond that, and further study polarization properties, the exchange Hamiltonian can explicitly account for phases of electron-hole exchange interaction~\cite{Ivchenko.Book.2005,tonin}
\begin{align}
\mathbf{H}_{\mathrm{exch}} &=\frac{1}{2}
\begin{bmatrix}
    \delta_{\mathrm{0}} & \delta_1  e^{-2i\theta_1} & 0 & 0  \\
    \delta_1 e^{ 2i\theta_1}  & \delta_{\mathrm{0}} & 0 & 0  \\
    0 & 0 & -\delta_{\mathrm{0}} & \delta_2 e^{ -2i\theta_2}  \\
    0 & 0 & \delta_2 e^{ 2i\theta_2} & -\delta_{\mathrm{0}}.
\end{bmatrix},
\label{Hexchrot}
\end{align}
where $\theta_1$ describes the rotation of polarization axis (around the growth axis $z$) by the angle $\theta_1$ with respect to the fixed axes, usually taken as crystal axis $[110]$. The eigenstates have the following form
\begin{align}
\ket*{BE_{1,2}}&=\left(\ket*{1}\pm e^{-2i\theta_1}\ket*{-1}\right)/\sqrt{2}\\
\ket*{DE_{1,2}}&=\left(\ket*{1}\pm e^{-2i\theta_2}\ket*{-1}\right)/\sqrt{2}
\end{align}
By comparing with atomistic calculations, we find that non-alloyed $C_{2v}$ system $2\theta_1 = \pi$, and 
$2\theta_2 = \frac{3}{2}\pi$ indicating $\pi/2$ phase difference between dark and bright manifolds.

While Eq.~\ref{Hexchrot} allows understanding rotation of quantum dot principal axes with respected to crystal axes, it still by definition predicts exactly vanishing polarization anisotropy,~\cite{tonin}
i.e., $C = \left(I_{max}-I_{min}\right)/\left(I_{max} + I_{min}\right) = \left(I_{BE2}-I_{BE1}\right)/\left(I_{BE2} + I_{BE1}\right)= 0$
since magnitudes of emission intensities of both bright excitons are exactly equal.

To go beyond this picture, and allow for better agreement with the experiment, one can account for valence-band mixing, which occurs in realistic quantum dots, due to shape asymmetry and lattice anisotropy, leading to an addmixture of light-hole component
\begin{equation}
\left|\pm \widetilde{1}\right>=\sqrt{1-\beta^2}\left|\pm\frac{3}{2},\mp\frac{1}{2}\right>+\beta e^{\pm 2i\psi}
\left|\mp\frac{1}{2},\mp\frac{1}{2}\right>,
\label{basis}
\end{equation}
where $\beta$ and $\psi$ are amplitude and phase of mixing with light-hole exciton $\left|\mp\frac{1}{2},\mp\frac{1}{2}\right>$, and tilde symbol was used to distinguish from the basis with no band-mixing included.
A Similar formula is used for dark states
\begin{equation}
\left|\pm \widetilde{2}\right>=\sqrt{1-\beta^2}\left|\pm\frac{3}{2},\pm\frac{1}{2}\right>+\beta e^{\pm 2i\psi}
\left|\mp\frac{1}{2},\pm\frac{1}{2}\right>.
\label{basis}
\end{equation}

Exchange Hamiltonian expressed in the basis $\ket*{\pm\widetilde{1}},\ket*{\pm\widetilde{2}}$ maintains the block diagonal structure of Eq.~\ref{Hexch}, and its eigenstates yet with tilded states replacing $\ket*{\pm 1}, \ket*{\pm 2}$.
$\beta$ is responsible for polarization anisotropy ~\cite{tonin,leger}
{\begin{equation}
C\left(\beta\right)=\frac{2\beta\sqrt{3\left(1-\beta^2\right)}}{3-2\beta^2},
\label{deg-pol}
\end{equation}}
Polarization anisotropy is naturally present in atomistic calculation, and for a $C_{2v}$ system it reaches 2.9\% with emission intensity for $\ket*{BE2}$ along $[110]$ larger then for $\ket*{BE1}$ along $[1\overline{1}0]$ direction.
Recasting atomistic $C_{2v}$ results on valence-band-mixed $4\times4$ Hamiltonian gives $\beta$ value of 2.5\%.
Moreover, for $C_{2v}$ quantum dots mixing phase $\psi=0$, and emission lines remain orthogonal.~\cite{tonin, tsitsishvili2017impact}

For alloyed systems, both $\beta$ and $\psi$ determine an angle between excitonic lines, which in principle can be different from 90$\degree$.
\begin{figure}
 \begin{center}
  \includegraphics[width=0.48\textwidth]{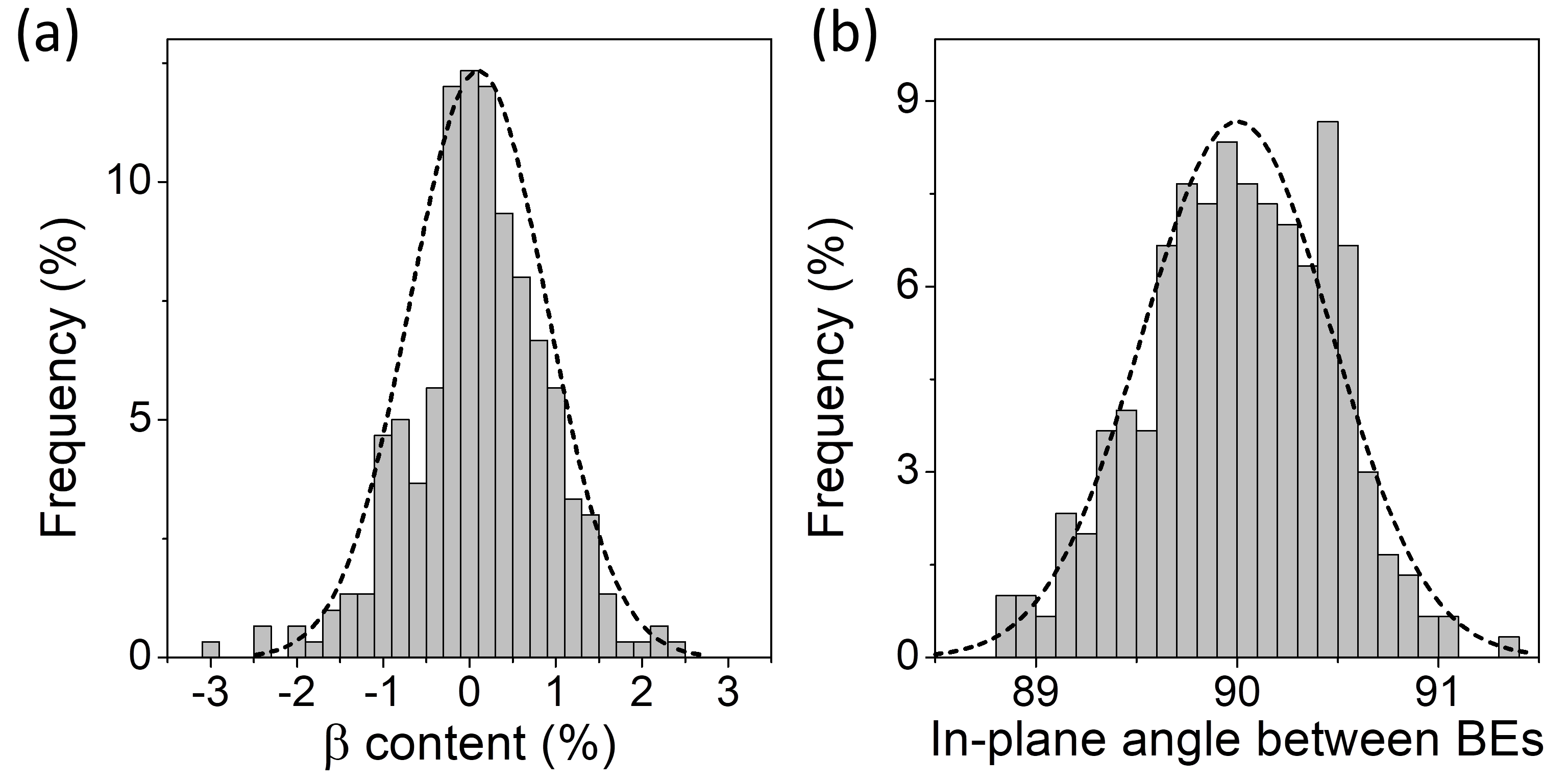}
  \end{center}
  \caption{Histogram of (a) $\beta$ light-hole content for ensemble of alloyed quantum dots, and (b) in-plane angle difference between lower (BE1) and higher (BE2) energy bright exciton states. 
  The dashed lines are plotted assuming normal distribution.
  Despite alloying $\beta$ content remains rather small, with the angle between bright excitons close to 90 degrees.
  See the text for more details.}
  \label{fig:beta_angle}
\end{figure}
Fig.~\ref{fig:beta_angle} (a) shows a histogram of $\beta$ values for alloyed systems studied in this paper, whitch appears to be normally distributed around zero light-hole content, with both positive and negative light-hole mixing amplitudes possible, and most of the cases corresponding to $|\beta|<1\%$. Moreover, as shown in Fig.~\ref{fig:beta_angle} (b) valence-band mixing allows for angle between polarizations of bright excitonic lines different from the exact 90$\degree$, however for the vast majority of cases this difference does not exceed 1$\degree$,
rendering Eq.~\ref{HBESatom} a valid approximation for the considered ensemble of quantum dots.

Valence band mixing, expressed via the non-zero $\beta$ content, has its effect on the dark manifold as well, since
\begin{equation}
\left|\pm \widetilde{2}\right>=\sqrt{1-\beta^2}\left|\pm\frac{3}{2},\pm\frac{1}{2}\right>+\beta e^{\pm 2i\psi}
\left|\mp\frac{1}{2},\pm\frac{1}{2}\right>,
\label{basis}
\end{equation}
and since matrix elements of the optical transition~\cite{Ivchenko.Book.2005,tsitsishvili2017impact} for the light-hole component are given as:
$M_{\mp\frac{1}{2},\pm\frac{1}{2}}\propto e_z$, dark excitons $\ket*{\pm \widetilde{2}}$ get optical activity due to hole band mixing, which is proportional to $\beta$, i.e., 
$|M_{\widetilde{\pm2}}|^2 = \beta^2 \left[1\mp \mathrm{cos}\;2(\theta_2 -2 \psi)\right]$.
For $C_{2v}$ system $\psi=0$,~\cite{tsitsishvili2017impact} and since from atomistic calculation: $2\theta_2=\frac{3}{2}\pi$, and $\delta_2>0$,  only one of dark exciton states (higher energy, DE2) is optically active, which allows the DE emission perpendicular to the
growth direction to be detected, and the other (lower energy, DE1) remains fully dark in agreement with experimental findings.~\cite{Smolenski.PRB.2012}

Moreover, group-theoretical arguments indicate that the optical dipole moment (matrix elements of the optical transition) for the dark exciton can be non-zero even for pure heavy-hole exciton, i.e., $\left|M_{DE2}\right|^2 \propto e_z$ even if $\beta=0$, whereas DE1 remains optically in-active, i.e., $\left|M_{DE1}\right|^2=0$.
Thus, the single-particle dipole matrix element, for $z$-polarized light, calculated in the basis of tight-binding states:
$|M^{z}|=|M^{z}_{e\uparrow,h\Uparrow}|=|M^{z}_{e\downarrow,h\Downarrow}|\propto |\beta M_{LH} + M_{HH}|$, and therefore could be interpreted as having light- and heavy-hole components.
\begin{figure}
 \begin{center}
  \includegraphics[width=0.48\textwidth]{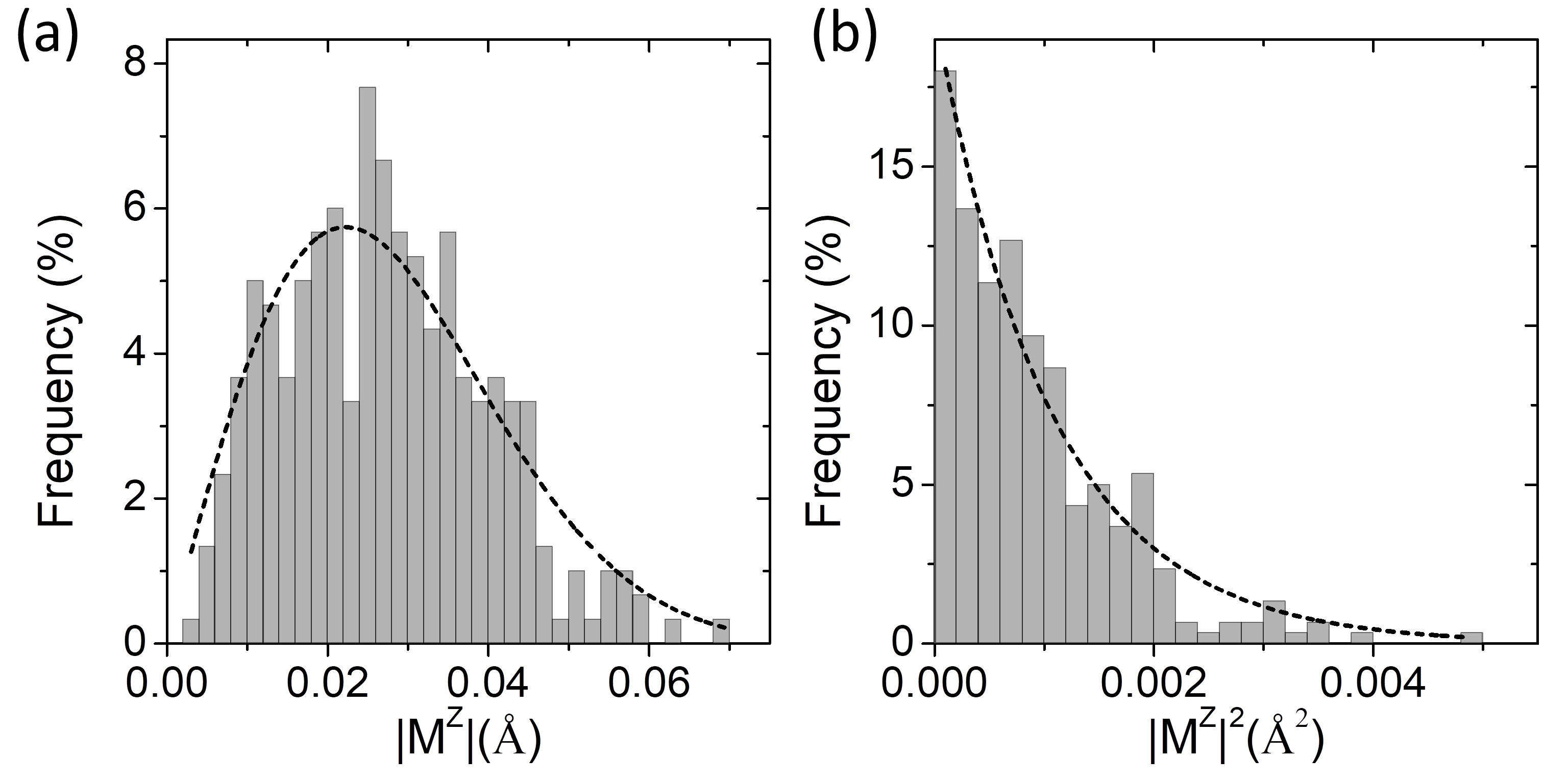}
  \end{center}
  \caption{Histogram of (a) modules of out-of-plane single-particle optical dipole moment $|M^{z}|=|M^{z}_{e\uparrow,h\Uparrow}|=|M^{z}_{e\downarrow,h\Downarrow}|$, and (b) its squared moduli. The dashed lines are calculated assuming probability distributions discussed in the text, suggesting the two-component character of the optical dipole moment.}
  \label{fig:M_stat}
\end{figure}
It not straightforward to perform such decomposition for atomistically obtained states. We have thus tried an approximate attempt by applying vertical electric field, aiming to use the field to reduce polarization anisotropy of bright exciton to zero, for which case, based on Eq.~\ref{deg-pol}, one can assume $\beta\sim0$. In such a situation $|M^{z}|$ should only have the heavy-hole component, which could be estimated in this way. By performing such analysis for the $C_{2v}$ system we found 
$M_{LH}=-0.042~\si{\angstrom}$ and $M_{HH}=0.051~\si{\angstrom}$.
\begin{figure*}
 \begin{center}
  \includegraphics[width=0.90\textwidth]{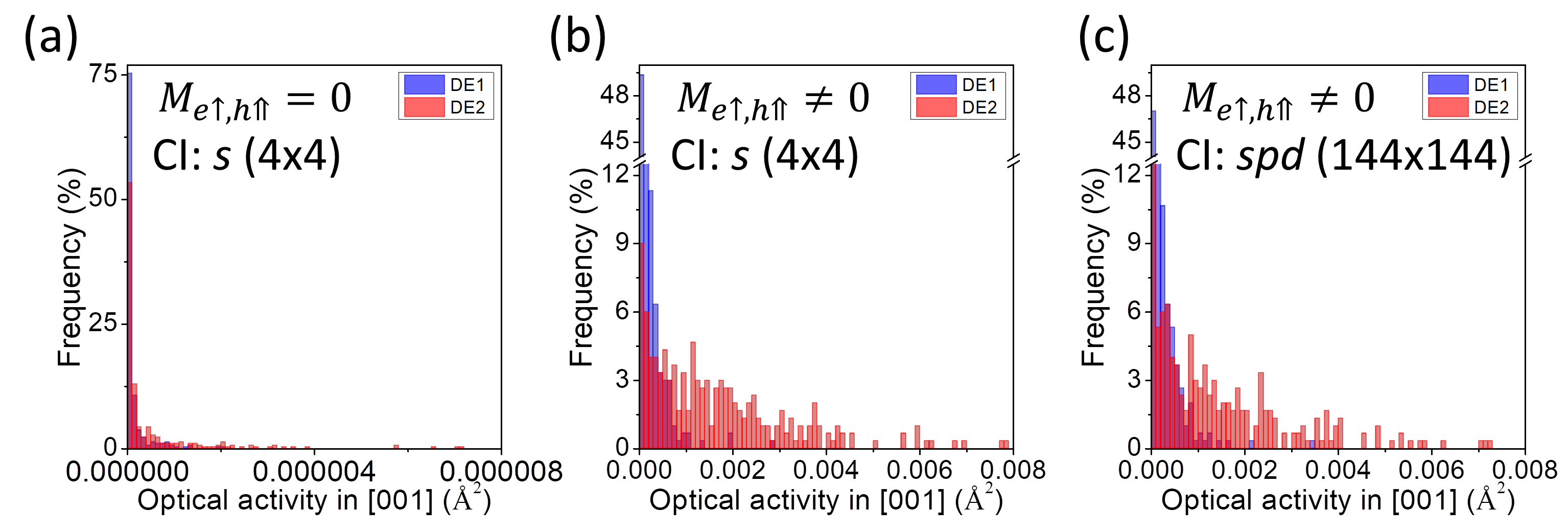}
  \end{center}
  \caption{Histogram of $z$-polarized (out-of-plane) optical spectra for lower (DE1) and higher (DE2) dark exciton states for cases with (a) optical dipole moment artificially set to zero ($M_{e\uparrow,h\Uparrow}=0$ and $M_{e\downarrow,h\Downarrow}=0$), and with (b) optical dipole moment included in the calculation, and (c) calculation performed in a basis including $s$, $p$, and $d$ shells of 144 excitonic configurations. 
  The $z$-polarized dark exciton activity thus stems from non-zero optical dipole moment (due to valence-band-mixing), whereas the exchange mixing plays a negligible role, and higher-shells play a rather small role as well.  See the text for more details.}
  \label{fig:de_z}
\end{figure*}
Therefore, both contributions can have comparable magnitudes, and can have opposite signs, thus partially cancel each other. Despite its approximated character, we have performed a similar study for 5 alloyed systems as well, and came to similar conclusions. To further analyze $|M^{z}|$ distributions in the considered ensemble of alloyed system, we show its histogram in Fig.~\ref{fig:M_stat} (a), and then $|M|^2$ histogram in Fig.~\ref{fig:M_stat} (b).
$|M^{z}|$ distribution can be well fit to $\chi$ distribution with two degrees (k=2) of freedom, which is mathematically equivalent to Rayleigh distribution/Wigner surmise discussed earlier 
(i.e., $\sim x\,\mathrm{exp}\left(-\frac{\pi}{4}x^2\right)$, with $x=|M^{z}|/\langle |M^{z}|\rangle$ shown as dotted line on Fig.~\ref{fig:M_stat} (a). Therefore, it is expected that, $|M^z|^2$ would be well described by $\chi^2$ distribution, again with two degrees of freedom, corresponding to the exponential distribution $\sim \mathrm{exp}\left(-x\right)$.
In other words, Fig.~\ref{fig:M_stat} does indirectly support two (light- and heavy-hole) component character of single-particle dipole moment $|M^z|$.
Aside from speculations regarding heavy- and light- hole contributions, since $M^z$ is a complex number, Rayleigh distribution is expected~\cite{,mehta2004random} when real and imaginary components are independently and identically (Gaussian) distributed (with equal variance and zero mean). In such case, the absolute value of the complex number is Rayleigh-distributed, as apparently shown in Fig.~\ref{fig:de_z}, thus supporting the two-component character of $M_z$ irrespective from its origin.

Finally, after an analysis of single-particle transition matrix elements, we can study many-body dark exciton optical spectra, shown for $z$-polarization in Fig.~\ref{fig:de_z}.
Should the single-particle optical dipole moments be artificially neglected, i.e., $|M_{e\uparrow,h\Uparrow}|=|M_{e\downarrow,h\Downarrow}|=0$, both dark excitons would not reveal any significant optical activity. The residual, weak activity [Fig.~\ref{fig:de_z} (a)] may be attributed to the dark-bright mixing term (due to exchange interaction) present in the atomistic calculation that will be discussed in the following part of the paper. In the case of $z$-polarized dark exciton emission exchange mixing apparently does not play any significant role. However, the inclusion of non-zero optical matrix element leads to substantial increase in $z$-polarized emission as seen in Fig.~\ref{fig:de_z} (b). This increase is particularly important for the higher-energy DE2 states, for which it can reach over 0.006 $\si{\angstrom}^2$, i.e., 
approximately 1/6000 fraction of bright exciton emission merely due to alloying.
DE1 states gains some optical activity as well, which is however much smaller, and what seem to be a property ``inherited'' from $C_{2v}$ system since
both amplitude and phase of valence band mixing i.e., $\beta$ (and apparently $\psi$) remain rather small in the alloyed system. Notably, further extension of the CI basis by inclusion of higher shells [Fig.~\ref{fig:de_z} (c)] does not seem to affect dark exciton optical spectra, indicating the dominant contribution is from the optical dipole moment, and not due to configuration mixing. 

In a phenomenological treatment, several authors further extend valence band mixing approach by accounting for a mixing term between heavy- and light-hole states with parallel spins:
\begin{align}
\left|\pm \dbtilde{1}\right>=&\sqrt{1-\beta^2-\gamma^2}\left|\pm\frac{3}{2},\mp\frac{1}{2}\right>\nonumber \\
&+\beta e^{\pm 2i\psi}\left|\mp\frac{1}{2},\mp\frac{1}{2}\right>
+\gamma e^{\pm 2i\xi}\left|\pm\frac{1}{2},\mp\frac{1}{2}\right>
\label{basis}
\end{align}
with a similar formula for $\ket*{\pm \dbtilde{2}}$ dark states, and where $\gamma$ and $\xi$ are the amplitude and the phase of mixing, and the double tilde symbol is used to distinguish from previously considered cases. This approach has been used to study the $z$-polarized emission of bright excitons,~\cite{tonin} with $\gamma$ parameter being far more difficult to extract from the experiment (or atomistic calculations) than $\beta$, and thus we will not aim for doing that. However, it was also shown recently ~\cite{germanis.PhysRevB.98.155303} that mixing $\gamma$ term induces off-diagonal Hamiltonian matrix elements (exchange integrals) responsible for the dark-bright exciton mixing:
\begin{align}
\begin{split}
\mathbf{H}_{\mathrm{exch}} &=\frac{1}{2}
\begin{bmatrix}
    \delta_{\mathrm{0}} & \delta_1  e^{-2i\theta_1} & \delta_{11} & \delta_{12}  \\
    \delta_1 e^{ 2i\theta_1}  & \delta_{\mathrm{0}} & -\delta_{12} & -\delta_{11}  \\
    \delta_{11}^* & -\delta_{12}^* & -\delta_{\mathrm{0}} & \delta_2 e^{-2i\theta_2}  \\
    \delta_{12}^* & -\delta_{11}^* & \delta_2 e^{2i\theta_2} & -\delta_{\mathrm{0}}
\end{bmatrix}\\
&=\frac{1}{2}
\begin{bmatrix}
    \mathbf{H}_{\mathrm{BB}} &  \mathbf{H}_{\mathrm{BD}}  \\
    \mathbf{H}_{\mathrm{DB}}  & \mathbf{H}_{\mathrm{DD}} 
\end{bmatrix}
\end{split}
\label{Hmixed}
\end{align}
where $\mathbf{H}_{\mathrm{DB}}$ denotes dark-bright coupling for brevity, whereas $\delta_{11}$, and $\delta_{12}$ are responsible for mixing of bright and dark configuration states differing by the electron and the hole spin state respectively, i.e., $\delta_{11}$ mixes $\ket*{\dbtilde{1}}$ with $\ket*{\dbtilde{2}}$, and $\delta_{12}$ mixes $\ket*{-\dbtilde{1}}=$ with $\ket*{-\dbtilde{2}}$.
With exception to phase factors, an identical Hamiltonian can be also derived in a phenomenological manner, by assuming the rotation of quantum-dot axes (angular-momentum quantization axes) from the ideal crystal axes.~\cite{don2016optical} Virtually identical Hamiltonian is also derived from atomistic calculations for $C_s$ quantum dots, where low shape symmetry induces mixing between bright and dark states.~\cite{Zielinski.PRB.2015}

\begin{figure}
 \begin{center}
  \includegraphics[width=0.48\textwidth]{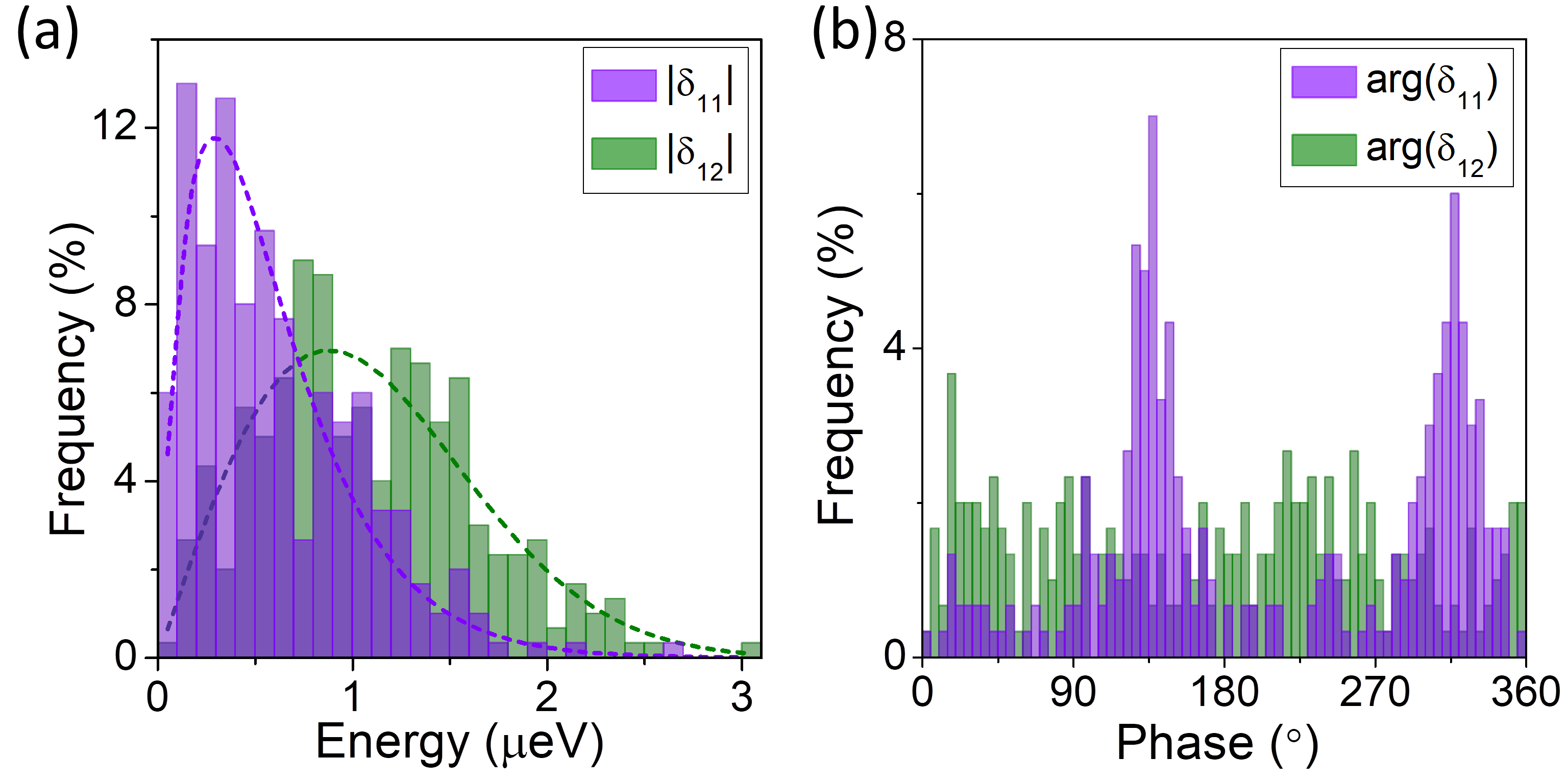}
  \end{center}
  \caption{Histogram of (a) moduli (absolute values) of $\delta_{11}$, and $\delta_{12}$ dark-bright mixing integrals, and (b) their arguments (phases). 
  See the text for details.}
  \label{fig:phase_coupling}
\end{figure}

\begin{figure}
 \begin{center}
  \includegraphics[width=0.48\textwidth]{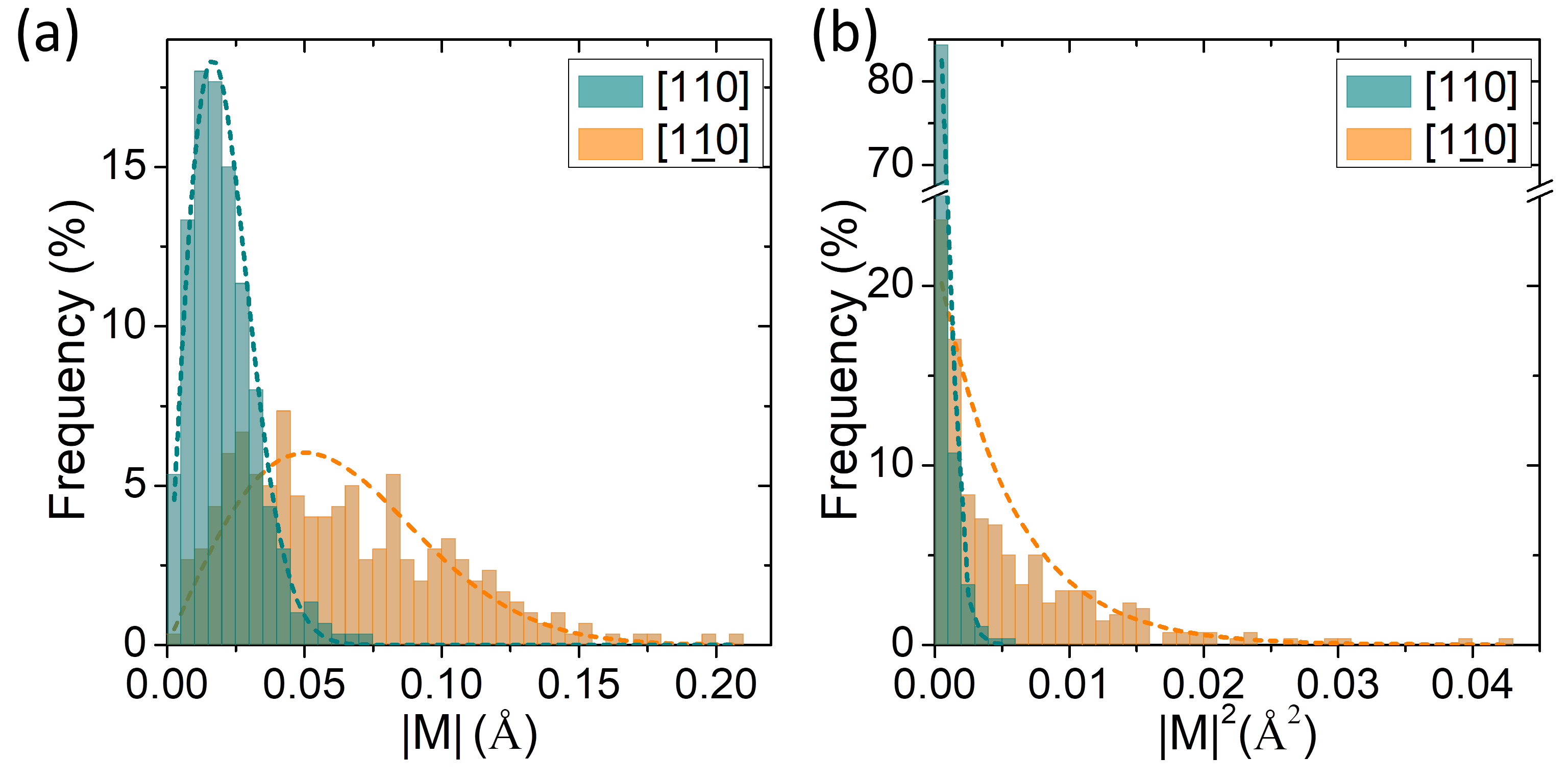}
  \end{center}
  \caption{Histograms of (a) moduli of in-plane single-particle optical dipole moments for light polarized in [110] and [1$\overline{1}$0] directions, i.e., $|M^{[110]}_{e\uparrow,h\Uparrow}|$ and $|M^{[1\overline{1}0]}_{e\downarrow,h\Downarrow}|$, (b) their squared moduli. The dashed lines are calculated assuming probability distributions discussed in the text. }
  \label{fig:M_inplane}
\end{figure}

Interestingly, we also found identical structure of (lowest four states) Hamiltonian from atomistic calculations for currently studied alloyed quantum dot despite their cylindrical symmetry, and lack of faceting.
Therefore, not only shape deformation can induce bright-dark exciton exchange coupling, but breaking the symmetry on the scale of local atomic arrangement can lead to the same effect.
Moreover, despite overall $C_1$ symmetry, the dark-bright coupling block derived from atomistic calculation, in a basis of tight-binding states $\ket*{e_\uparrow,h_\Uparrow},\ket*{e_\downarrow,h_\Downarrow}|$ is given as:
\begin{align}
\mathbf{H}_{\mathrm{DB}} &=
\begin{bmatrix}
    \delta_{11} & \delta_{12}  \\
    -\delta_{12}^* & -\delta_{11}^*
\end{bmatrix},
\label{HmixedTB}
\end{align}
where the only difference from phenomenological Hamiltonian Eq.~\ref{HmixedTB} (expressed in double-tilde basis) is related to complex conjugations in the second row.

\section{Dark exciton in-plane optical activity}
\label{section:dark}

To further study the role of dark-bright exciton mixing, in Fig.~\ref{fig:phase_coupling} (a) we show histograms of $\delta_{11}$ and $\delta_{12}$.
The absolute value of $\delta_{12}$ can be quite well fit to $x\,\mathrm{exp}\left(-x^2\right)$, which is again equivalent to $\chi$ distribution with k=2, suggesting two-component character of $\delta_{12}$ integral. This is consistent with Ref.~\cite{don2016optical}, where it was interpreted in terms of two effective tilt angles (one related to the electron, and second to the hole) leading to the dark-bright mixing. This is also coherent with approach of Ref.~\cite{germanis.PhysRevB.98.155303} where $\delta_{12}$ is induced by the presence of both $\beta$ and $\gamma$ mixing terms. In the treatment employed in Ref.~\cite{germanis.PhysRevB.98.155303} $\delta_{11}$ depends on $\gamma$ only. One can thus expect its statistical distribution to differ from that found for $\delta_{12}$, and in fact this is visible in Fig.~\ref{fig:phase_coupling} (a), where one can fit $\delta_{11}$ to $x\,\mathrm{exp}\left(-x\right)$, resembling Gamma distribution with k=1.
Whereas this sort of analysis is speculative, $\delta_{11}$ undoubtedly varies from ~0 to 2.7~$\mu$eV with mean (average) of 0.5~$\mu$eV and $\delta_{12}$ varyies from 0 to to 3.05~$\mu$eV with the mean of 1~$\mu$eV. 
Therefore, $\delta_{11}$ and $\delta_{12}$ have magnitudes comparable with $\delta_1$, the dark exciton splitting, and thus \textit{a priori} cannot be neglected when considering excitonic fine structure of an alloyed system.
Moreover, both BES and DES will be renormalized by the presence of dark-bright coupling~\cite{germanis.PhysRevB.98.155303,don2016optical}. However, since the correction is equal to $\delta_{11}\delta_{12}/\delta_0$, the effect of dark-bright mixing on both BES and DES is negligible for the system considered in this work, as we have $\delta_0 \gg \delta_{11,12}$.

For $C_{2v}$ quantum dots, by symmetry, both dark excitons are strictly optically inactive, when considering the in-plane polarization.
For $C_1$ quantum dot however, there are two channels, through which dark excitons can gain in-plane optical activity (see the Appendix~C for more details). 
One of them is related to the dark-bright exchange interaction, which was already mentioned above, and in which the dark exciton can gain optical activity from the admixture of bright configurations.
The second channel is related to a non-vanishing dipole moment for in-plane ($x$,$y$) polarized light
$M^{x,y}_{e\uparrow,h\Uparrow}\neq 0$, and 
$M^{x,y}_{e\downarrow,h\Downarrow}\neq 0$, similarly to the $M^{z}$ dipole moment discussed earlier.

Fig.~\ref{fig:M_inplane} shows histograms of in-plane dipole moment matrix elements along two experimentally relevant crystal axes, [110] and [1$\overline{1}$0]. 
Importantly, $|M|^{[1\overline{1}0]}$ has substantially larger magnitude than $|M|^{1\overline{1}0]}$, thus despite alloying and cylindrical shape of quantum dots in the ensemble,
the lattice anisotropy still plays an important role with
average $\langle|M|^{[1\overline{1}0]}\rangle=0.06304~\si{\angstrom}$ about 3 times larger than $\langle|M|^{[110]}\rangle= 0.02061$. 

\begin{figure}
 \begin{center}
  \includegraphics[width=0.48\textwidth]{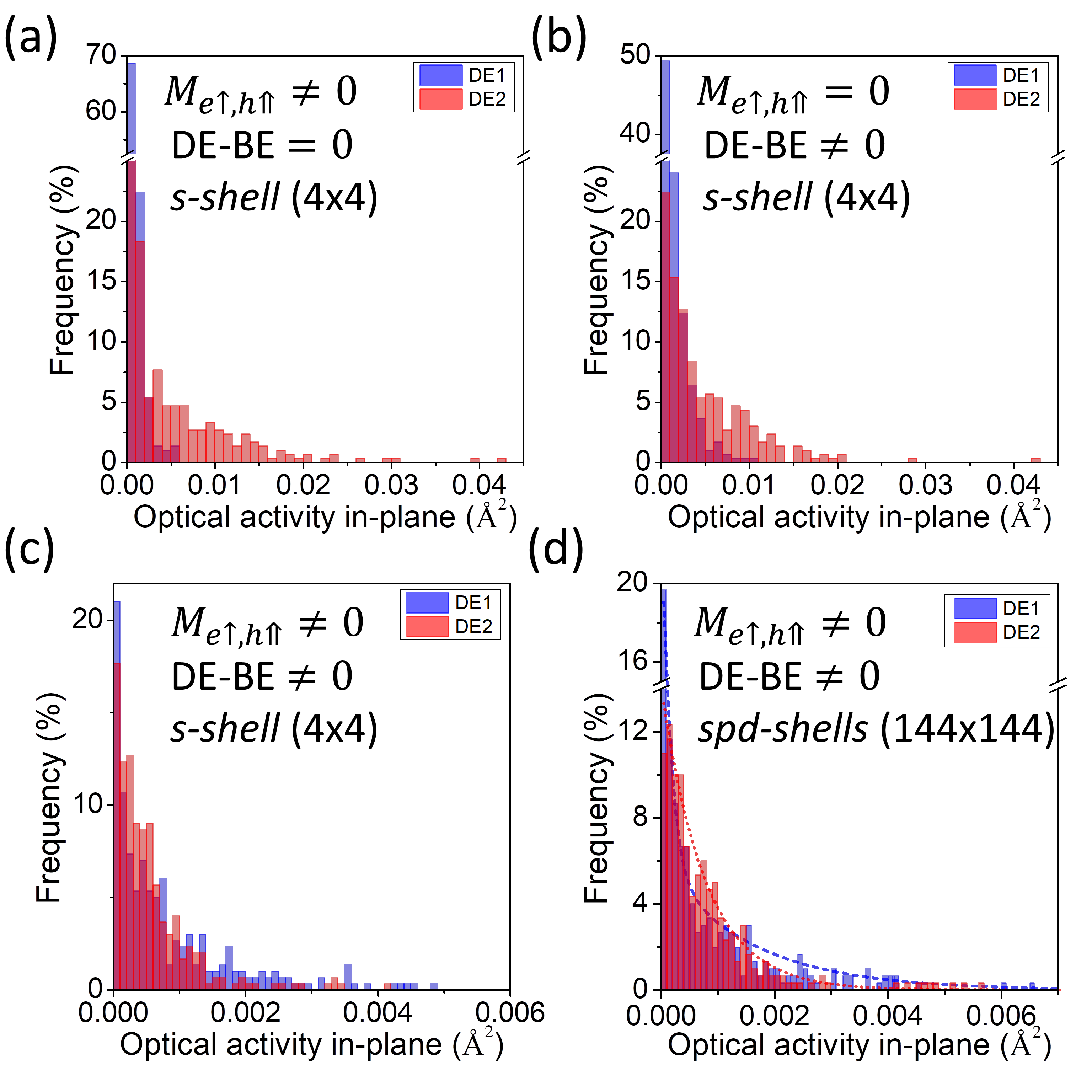}  \end{center}
  \caption{Histogram of in-plane optical spectra (emission intensity) for lower (DE1) and higher (DE2) dark exciton states with (a) 
  optical single-particle dipole moment included in the calculation, but exchange dark-bright mixing terms neglected ($\delta_{11}=\delta_{12}=0$) in the Hamiltonian,
  (b) optical dipole moment artificially set to zero ($|M_{e\uparrow,h\Uparrow}=0$ and $M_{e\downarrow,h\Downarrow}=0$), but exchange mixing accounted for, (c) both contributions included, and (d) as in (c) but with calculation performed in a basis including $s$, $p$, and $d$ shells of 144 excitonic configurations. 
  Note different scales, as well as presence of breaks on vertical axes.
  Apparently, both non-zero optical dipole moments, and exchange mixing play equally important roles, and none of these contributions can be neglected, as it would overestimate the optical activity.
  See the text for more details.}
  \label{fig:DE_inplane}
\end{figure}

We note as well that $|M|^{[110]}$ is larger than the one for the $z$-polarized case, i.e., $\langle|M|^{001]}\equiv\langle|M|^{z}\rangle$ (shown earlier in Fig.~\ref{fig:M_stat})
This may be understood since quantum-dot lateral dimensions are substantially larger than the in growth ($z$) directions. 
However, similarly to $|M^{z}|$, $|M|^{110}]$ can be very well approximated by $\chi$ distribution with k=2, and with  $|M|^2$ [Fig.~\ref{fig:DE_inplane} (b)] showing a $\chi^2$ dependence, with k=2, i.e., the exponential distribution.

Since in-plane single-particle dipole moment matrix elements are non-zero, as are the dark-bright exchange mixing terms in the Hamiltonian, it is instructive to study how these terms individually affect many-body dark exciton optical spectra. To this end, in Fig.~\ref{fig:DE_inplane} (a) we show the results with dark-bright mixing artificially neglected, and with single-particle dipole moment being the only term inducing dark exciton in-plane optical activity.
In this case DE2 has much larger optical activity than DE1, similarly to previously considered out-of-plane emission. Alternatively, exchange mixing can be artificially neglected and optical dipole moments accounted for, as in Fig.~\ref{fig:DE_inplane} (b).
Such a situation produces very similar optical spectra to the first case, with DE2 having substantially larger optical activity then DE1, despite a different mechanism of luminescence.

\begin{figure}
 \begin{center}
  \includegraphics[width=0.48\textwidth]{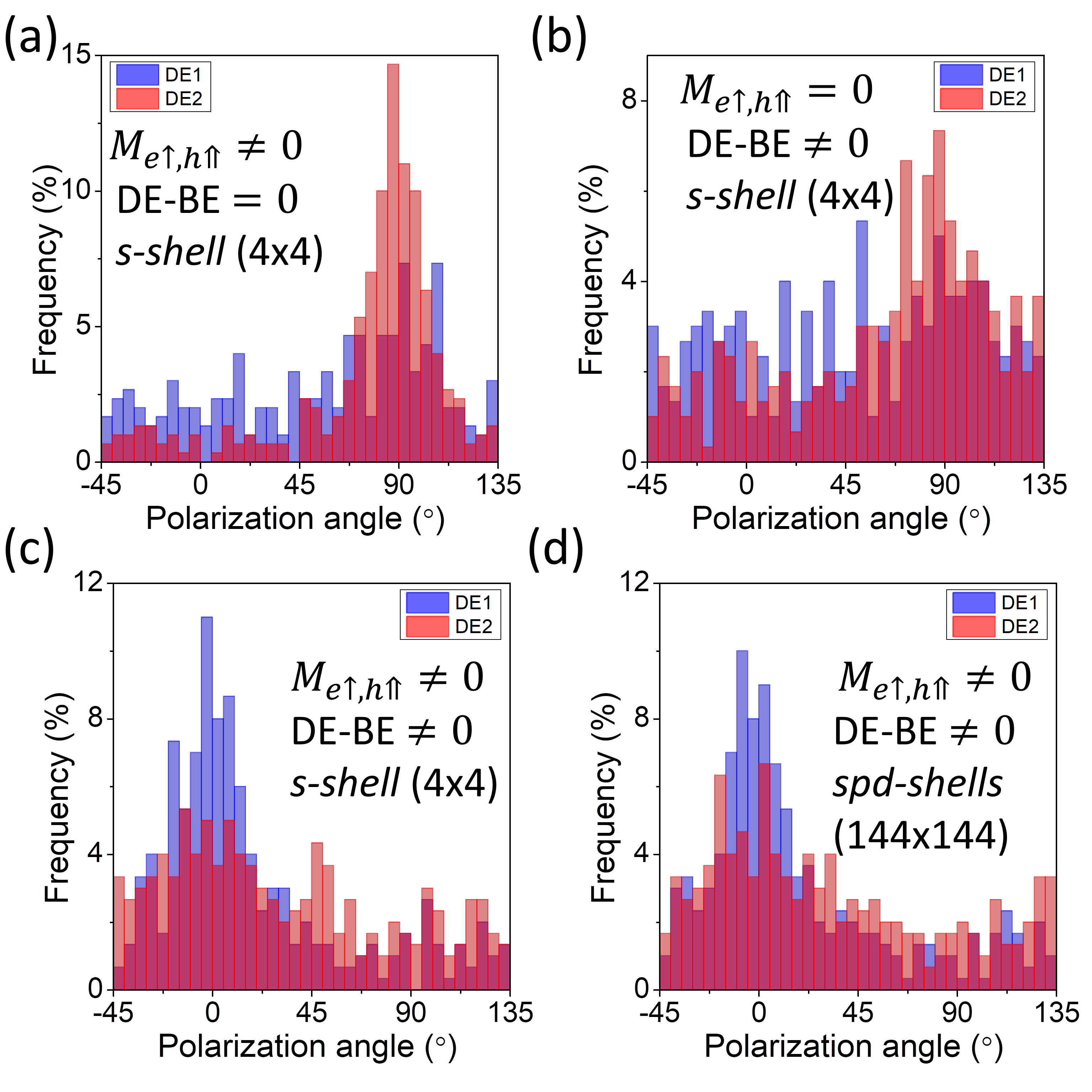}  \end{center}
  \caption{Histogram of in-plane polarization angle for lower (DE1) and higher (DE2) dark exciton states with (a)
  optical single-particle dipole moment included in the calculation, but exchange dark-bright mixing terms neglected ($\delta_{11}=\delta_{12}=0$) in the Hamiltonian,
  (b) optical dipole moment artificially set to zero ($|M_{e\uparrow,h\Uparrow}=0$ and $M_{e\downarrow,h\Downarrow}=0$), but exchange mixing accounted for, (c) both contributions included, and (d) as in (c) but with calculation performed in a basis including $s$, $p$, and $d$ shells of 144 excitonic configurations. 
  Again, both non-zero optical dipole moments, and exchange mixing play equally important roles, and none of these contributions can be neglected.
  See the text for more details.}
  \label{fig:DE_inplane_angle}
\end{figure}

Interestingly, when both terms are included [Fig.~\ref{fig:DE_inplane} (c)], they appear to partially compensate each other, with the dark exciton optical activity on average an order of magnitude smaller compared to (a) and (b) cases, thus comparable with the out-of-plane emission. A notable difference to z-polarized emission can however be observed, since both DE1 and DE2 have (on average) similar optical activity. 

Therefore, reduction of symmetry due to alloying must be accounted for in both transition dipole moment and exchange matrix elements on equal footing.
Failing to do so leads to a serious overestimation of the dark exciton in-plane optical activity. This could be understood in the spirit of Ref.~\cite{don2016optical} where the reduction of symmetry is related to rotation of quantum dot axes. This new quantization axis, or rotation of angular momenta, must be included simultaneously in the calculation of single-particle dipole moment, and two-body Coulomb and exchange integrals.

Moreover, we find that for $C_1$ alloyed quantum dots it is impossible to fit atomistic spectra to the effective $4\times4$ Hamiltonian (formally resembling Eq.~\ref{Hmixed}), yet with idealized matrix element of the
optical transition (i.e., vanishing in-plane optical dipole moments), and with $\delta_{11}$, $\delta_{12}$ being the only factors responsible for the dark-exciton optical activity.
Such procedure was indeed possible for $C_s$ systems,~\cite{Zielinski.PRB.2015} however for $C_1$, since both dipole moments and Coulomb matrix are affected by allowing, one would need to include two additional fitting parameters in the Hamiltonian ($\delta_{21}$ and $\delta_{22}$ integrals) to further differentiate coupling of bright and dark states, what does not have a formal justification. Thus, we again conclude that calculation of dipole moments and Coulomb matrix elements in alloyed system must be performed on equal footing, unless some other factor (speculatively large quantum dot shape deformation/anisotropy) dominates over alloying.

Further inclusion of higher shells beyond the $4\times4$ CI treatment does not change the overall picture [Fig.~\ref{fig:DE_inplane} (d)] considerably.
In this case, despite complicated origin of dark exciton activity (dipole moments, exchange mixing, and CI)
the frequency distribution of optical activity can be quite well fit with a single exponent for DE2 states, and with two exponents for DE1. 
Interestingly, optical activity larger than 0.002~$\si{\angstrom}^2$ seems to be dominated by that coming from DE1 states. So for several samples, the combined effect of non-zero dipole moments, and dark-bright mixing leads noticeably to more pronounce emission from the lower energy dark exciton state, although this is a rather subtle effect. Such activity would be in agreement with experimental results,~\cite{PhysRevX.5.011009}, although here we do not aim at comparison with a particular experiment. 
Due to alloy randomness, the dark exciton in-plane activity reaches 1/6000 fraction of the bright exciton activity, with the average of 1/40000, thus considerably smaller that observed when a shape distortion~\cite{Zielinski.PRB.2015} is the source of dark excitonic optical activity.
Nevertheless, we can conclude that a mere reduction of overall symmetry due to alloy randomness is able to trigger non-negligible in-plane polarized optical activity of dark excitons in self-assembled InGaAs quantum dots.

Finally, it is worth to conduct a similar analysis for polarization angle of the dark exciton in-plane emission, as shown in Fig.~\ref{fig:DE_inplane_angle}.
With either dark-bright exchange mixing artificially neglected [Fig.~\ref{fig:DE_inplane_angle} (a)] or 
dipole moments set to zero [Fig.~\ref{fig:DE_inplane_angle} (b)], the angle of polarization of both dark excitons tends to prefer 90$\degree$ rotation from [110], i.e., polarization along [1$\overline{1}$0] direction.
This is particularly pronounced for DE2 states. Therefore, with either of contributions  accounted for, whilst the other being neglected, the in-plane spectra of dark excitons somewhat resemble the out-of-plane ($z$) polarized spectra with DE2 having dominant optical activity.

However, the situation is apparently different with both contributions simultaneously included, as shown in Fig.~\ref{fig:DE_inplane_angle} (c), where DE1 (in particular) and DE2 (to a lesser degree) preferable polarization is along [110] (polarization angle 0$\degree$).
This is somewhat opposite to experimental observation, where the dominant emission is from the lower dark exciton state, following polarization from the lower energy bright exciton state. However, a caution must be exercised since experiments are unavoidably performed on non-ideal, low shape symmetry QDs.
Our atomistic results~\cite{Zielinski.PRB.2015} for a quantum dot with a facet, and thus broken shape symmetry, indeed indicated a very good agreement with a particular experiment~\cite{PhysRevX.5.011009,PhysRevB.92.201201}. Thus, a deviation from cylindrical base very likely must be included when one aims for a comparison with a particular experimentally grown system, and here we focus on alloy randomness effect only.

\begin{figure}
 \begin{center}
  \includegraphics[width=0.24\textwidth]{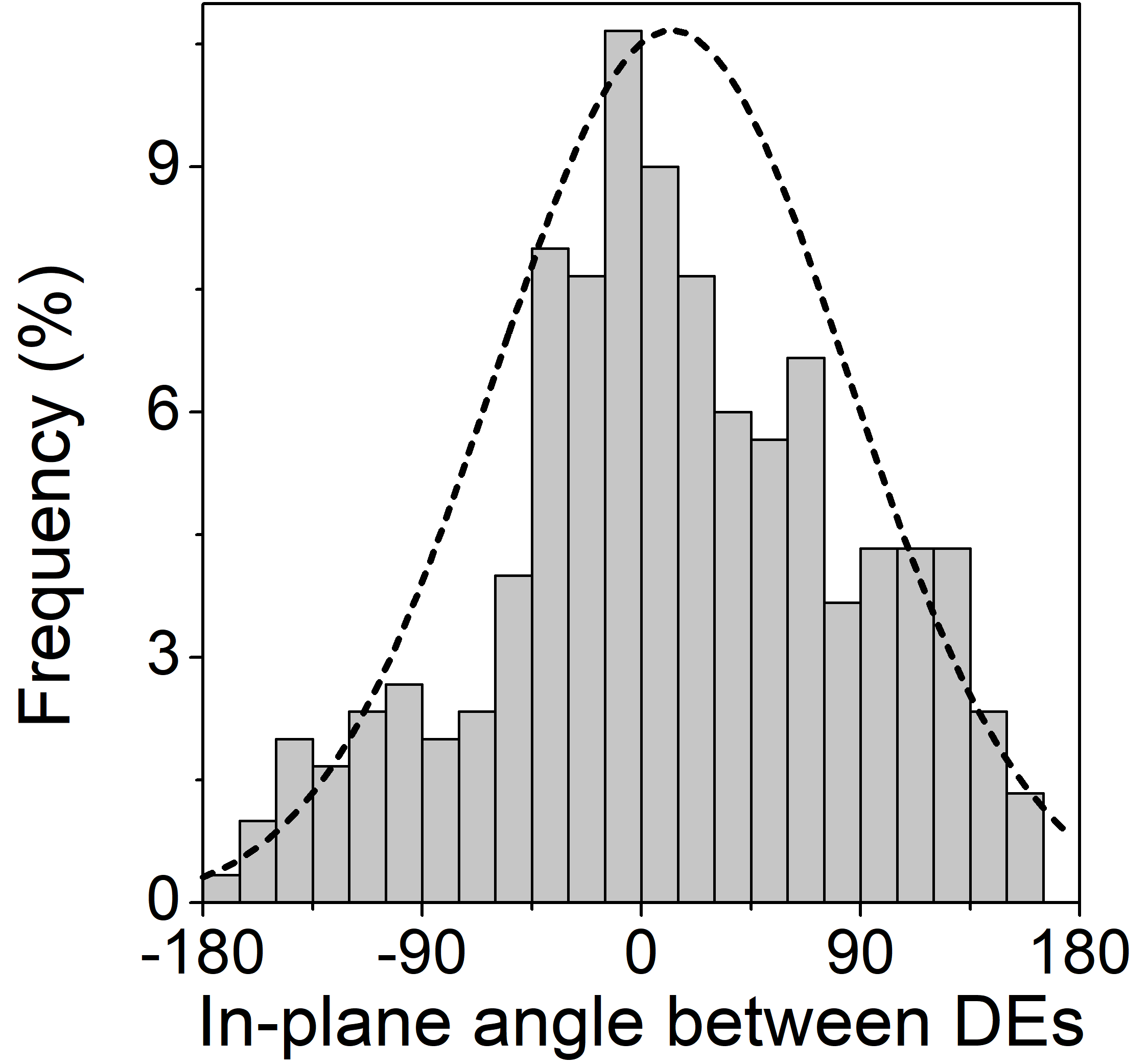}  \end{center}
  \caption{Histogram of in-plane angle difference between the lower (DE1) and the higher (DE2) energy dark exciton states.
  The dashed line is plotted assuming a normal distribution. Contrary to bright excitons the dark excitonic lines tend to have the same polarization angle, yet with a very broad distribution. 
  See the text for more details. }
  \label{fig:de_inplane_angle_diff}
\end{figure}

Since both DE1 and DE2 appear to have their polarization preferably along the [110] direction, it is instructive to show the histogram of angles between polarizations of DE1 and DE2 states (for their in-plane emission) as shown in Fig.~\ref{fig:de_inplane_angle_diff}. Contrary to bright excitons [shown earlier in Fig.~\ref{fig:besdeshist} (b)], where the polarization directions strongly tend to be orthogonal (i.e., the angle between polarizations close to 90$\degree$), the dark exciton states tend to be polarized along the same direction, however due to alloy randomness virtually any angle between these two emission lines is possible. Again, accounting for shape elongation or faceting in a particular experimental realization, together with alloy randomness could very likely change these statistics.

Moreover, there is still little experimental data regarding dark exciton spectra, since measurements are usually performed on a single quantum-dot sample only. Our calculation indicates, that for alloyed quantum dots one can cherry-pick a sample with DE1 polarized along [110] or [1$\overline{1}$0], with DE1 being the dominant line, or vice-versa. Therefore, more experimental research on the subject is needed.

\section{Radiative lifetimes}
\label{section:lifetime}

\begin{figure}
 \begin{center}
  \includegraphics[width=0.24\textwidth]{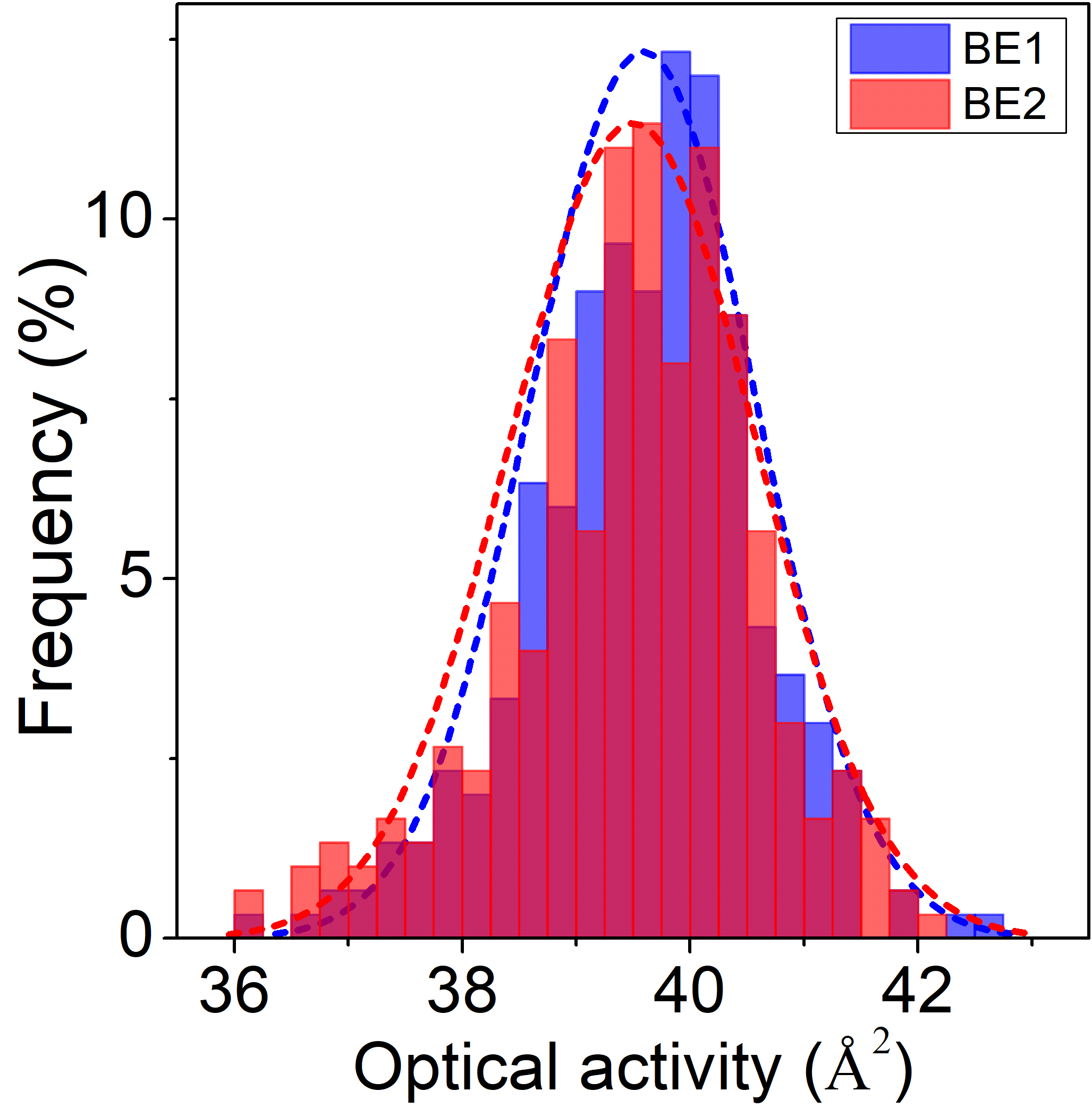}  \end{center} 
  \caption{Histograms of bright exciton states BE1 and BE2 optical activities, showing very similar and relatively narrow distributions. 
  Dashed lines are plotted assuming a normal distribution.}
  \label{fig:BEopt}
\end{figure}

\begin{figure}
 \begin{center}
  \includegraphics[width=0.48\textwidth]{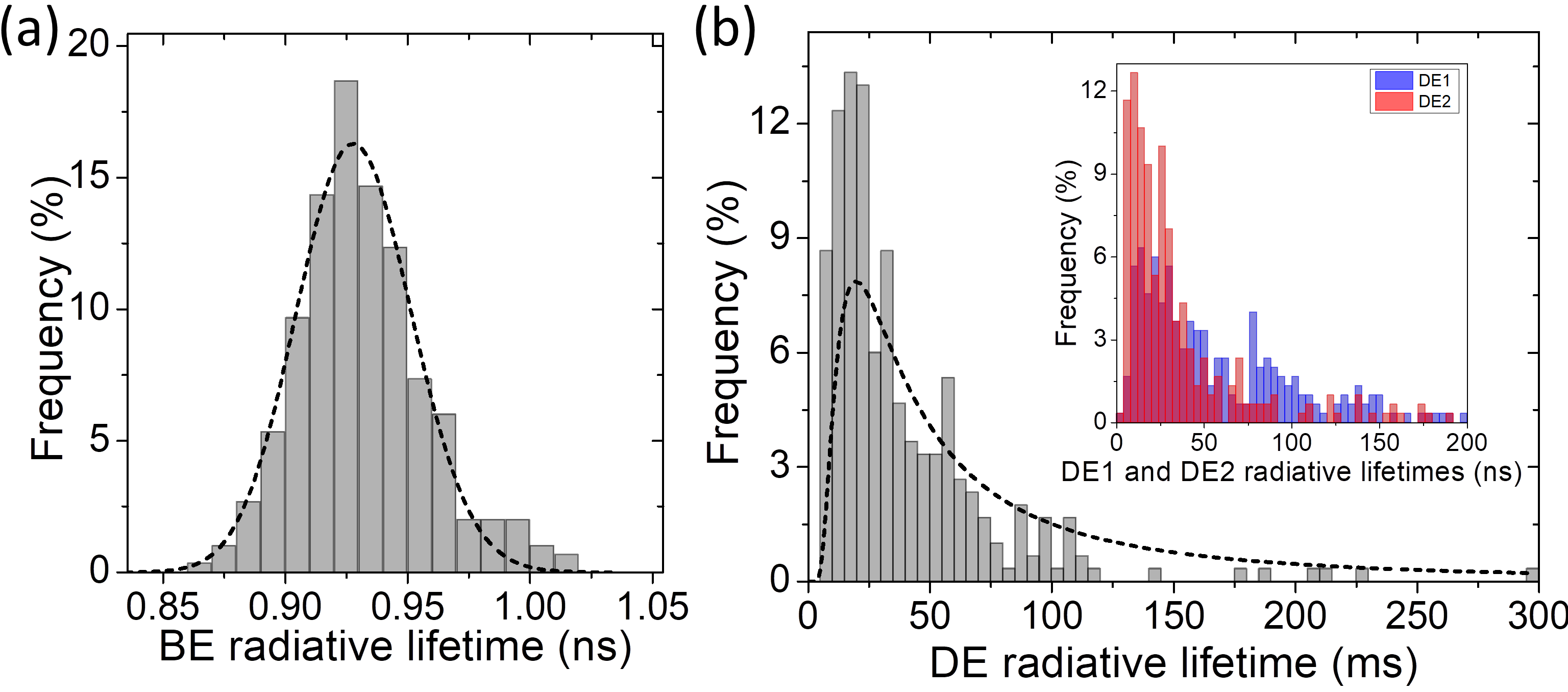} \end{center}
  \caption{Histogram of (a) bright exciton radiative lifetime (b) dark exciton radiative lifetime for the ensemble of alloyed quantum dots studied in this work.
  The inset shows lifetime of DE1 and DE2 states separately.
  The dashed lines are plotted assuming a normal distribution for the bright exciton, and inverse-gamma distribution for the dark exciton. See the text for more details. }
  \label{fig:lifetimes}
\end{figure}

In Section~\ref{section:dark} we have studied the dark exciton optical activity, therefore for comparison 
Fig.~\ref{fig:BEopt} shows histogram of both bright exciton states optical activities which are apparently normally distributed around approximately 39.5~$\si{\angstrom}^2$, corresponding
to mean optical dipole moment of 0.63~nm, with rather small variance corresponding to dipole moments from 0.6 to 0.66 nm,
thus very close to typical experimental results (e.g., 0.59~nm determined in Ref.~\cite{karrai2003optical}).

Finally, Fig.~\ref{fig:lifetimes} shows radiative lifetimes of bright and dark excitons calculated with use of Eq.~\ref{Eq:lifetime} and Eq.~\ref{Eq:lifetime2}.
Reported lifetimes are similar for both bright excitons, hence $\tau_{BE}\approx\tau_{BE1}\approx\tau_{BE2}$ with again apparently normal-like distribution and with the mean value of 0.93~nm, thus comparable to experimentally and theoretically reported values of approximately 1~ns, thus again close to typical quantum dot results.~\cite{karrai2003optical}
Despite its simplicity such result needs a comments, since lifetime is inversely proportional to optical activity (Eq.~\ref{Eq:intensity} and Eq.~\ref{Eq:lifetime}), thus the inverse of normal distribution on  Fig.~\ref{fig:BEopt} will lead to (in principle quite complicated) inverse Gaussian distribution~\cite{chhikara1988inverse}
$\propto\sqrt{\frac{\lambda}{x^3}}exp(-\frac{\lambda\left(x-\mu\right)^2}{2x\mu^2})$.
However, since the bright exciton optical activity has much larger mean value (approximately 40~$\si{\angstrom}^2$) than its standard deviation (of approximately 1~$\si{\angstrom}^2$), thus the inverse Gaussian distribution resembles regular Gaussian distribution. Fit to this dependence is shown on Fig.~\ref{fig:lifetimes} (a) as black dashed line, with $\lambda=1541$ and $\mu=0.9284$~ns, and
such large $\lambda$ being the reason~\cite{chhikara1988inverse} why the inverse Gaussian distribution appears very similar to the regular normal distribution.

The distribution of dark exciton lifetimes is different, with average (mean) $\tau_{DE}$ value equal to 39 ms, and different distribution. Since emission intensity (Eq.~\ref{Eq:intensity}) and lifetimes (Eq.~\ref{Eq:lifetime}) are inversely proportional, the weak dark exciton optical intensity, as shown earlier in Eq.~\ref{fig:de_z} and Eq.~\ref{fig:DE_inplane_angle}, corresponds to long lifetimes.
Moreover since dark exciton optical activity resembles the exponential $\mathrm{exp}(-x/\expval{x})$ distribution, mathematically equivalent to Gamma distribution~\cite{papoulis1965random} with k=1, then the inverse 
should have the distribution in the following form 
$\frac{1}{x^2}\mathrm{exp}\left(-\expval{x}/x\right)$ shown as a dashed line in Fig.~\ref{fig:lifetimes} (b). Despite it over-simplicity, the inverse gamma distribution qualitatively well describes the distribution of dark exciton lifetime with no quantum dots with lifetime below 6 ms and with a long tail of quantum dots with lifetimes reaching up to 300 ms. Such extremely long lifetime should be taken with great care, since we consider only the radiative contribution, obtained within tight-binding Hamiltonian and with other approximations included.~\cite{zielinski-prb09,sheng2012electronic}
We should also note that contrary to bright excitons (and as a direct result of differences in their optical spectra), the dark exciton states DE1 and DE2 reveal some difference of radiative lifetimes [inset of Fig.~\ref{fig:lifetimes} (b)]. 
Namely, the mean lifetime of DE2 state is equal to 41~ms, whereas for DE1, including the lowest 95\% cases, it is equal to 76~ms
with a long tail [not shown in Fig.~\ref{fig:lifetimes} (b)] extending up to seconds (with average of 390~ms of the 5\% highest-value subset), thus corresponding to virtually optically non-active state, where other processes would likely dominate over radiative recombination.
Since $\tau_{DE2}\gg\tau_{DE1}$, the overall DE radiative lifetime is limited by recombination through emission from DE2. Additionally, generally one observes that $\tau_{DE}\approx\tau_{DE2}$, which is consistent with the above-mentioned finding that in an ensemble of quantum dots considered in this work higher energy dark exciton state has (on average) larger optical activity for both out-of-plane and in-plane polarizations, although a sizable number of quantum dots has dominant emission through DE1 state, and short DE1 lifetime, with a strong variation within the ensemble.
 
\section{Summary}
To summarize, changes in the local atomic arrangement in an alloyed self-assembled quantum dot can trigger not only substantial fluctuation is BES and DES, but can also lead to a non-negligible in-plane optical activity of dark excitons.
Whereas the out-of-plane emission from dark exciton states is possible for higher-symmetry $C_{2v}$ quantum dots, and it origins from the valence-band mixing and lattice anisotropy, the in-plane emission in $C_{2v}$ systems is forbidden by symmetry. Contrarily, in an alloyed $C_{1}$ system dark excitons can emit in-plane polarized light. There are two contributions to this luminescence. One is related to non-vanishing matrix elements of the optical transition dipole moment, and the second is related to exchange mixing of bright and dark configurations in the configuration-interaction Hamiltonian.
Dark exciton optical spectra can significantly vary in the ensemble, from nanostructures having virtually vanishing oscillator strength, to quantum dots with a substantial 1/6000 fraction of the bright exciton intensity.
Our results indicate that, apart from shape-elongation and presence of facets, alloying must be accounted for in accurate modeling of quantum dot systems used a as building blocks of novel quantum devices. 

\acknowledgments
The author acknowledges the support from the Polish National Science Centre based on Decision No. 2018/31/B/ST3/01415.
The author would like to thank Michał Gawłeczyk for reviewing the manuscript and valuable comments.

\appendix

\section{BES and DES for s-shell only}
\begin{figure}
 \begin{center}
  \includegraphics[width=0.48\textwidth]{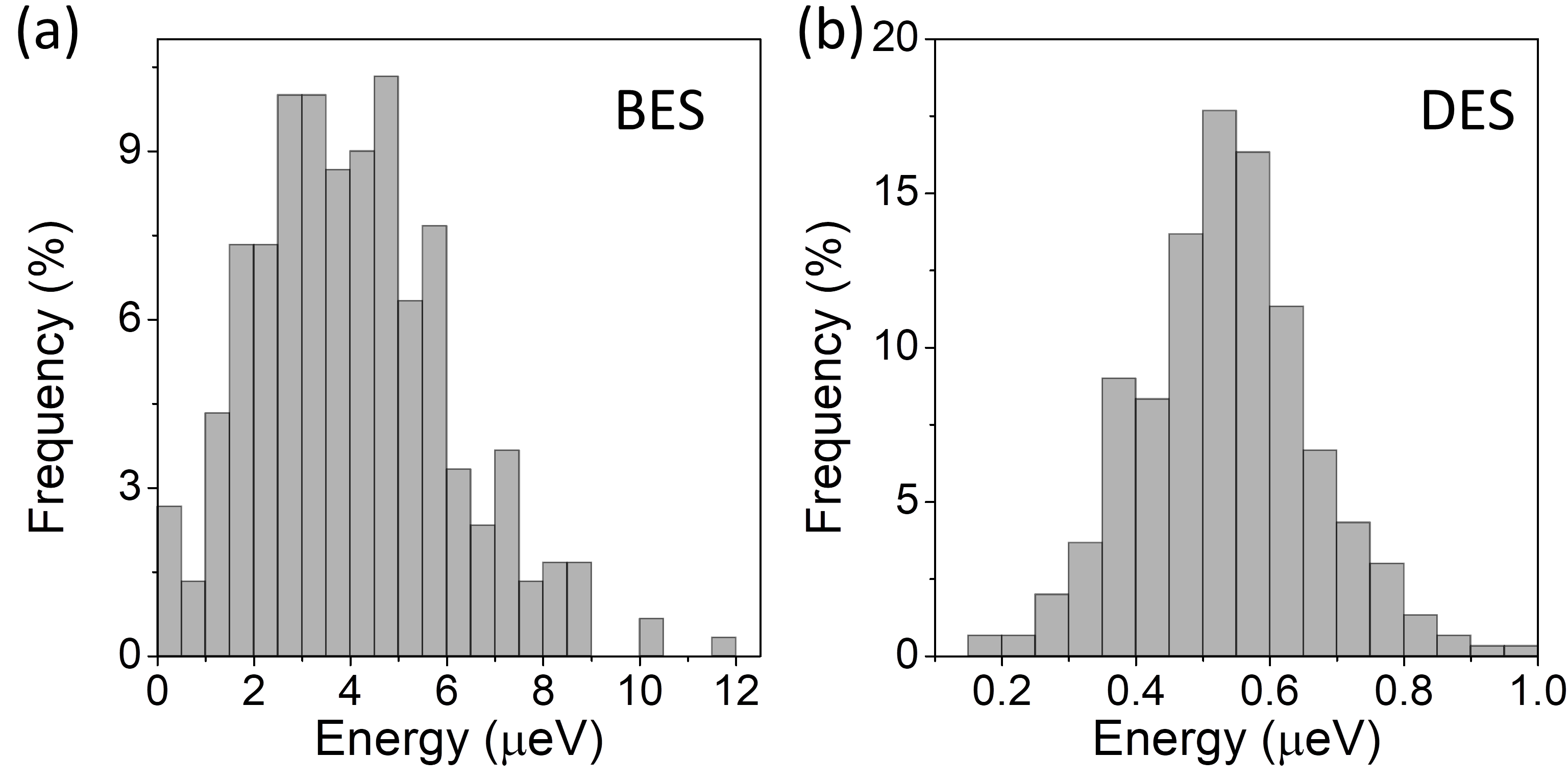}
  \end{center}
  \caption{
  Histograms of (a) the bright exciton splitting, and (b) dark exciton splitting calculated when accounting for the $s$ shell only in the configuration-interaction procedure.}
  \label{fig:app_s}
\end{figure}

Whereas Fig.~\ref{fig:splittings} in the main text shows BES and DES histograms for CI treatment including $s$, $p$, and $d$ shells, for comparison Fig.~\ref{fig:app_s} shows the (a) BES and (b) DES calculated when accounting for the s-shell only. In other words, Fig.~\ref{fig:app_s} (a) and (b) shows histograms of $\delta_1$ and $\delta_2$ exchange integrals, using notation of Eq.~\ref{Hexch}.

\section{Phases of dark and bright exchange integrals}
\begin{figure}
 \begin{center}
  \includegraphics[width=0.48\textwidth]{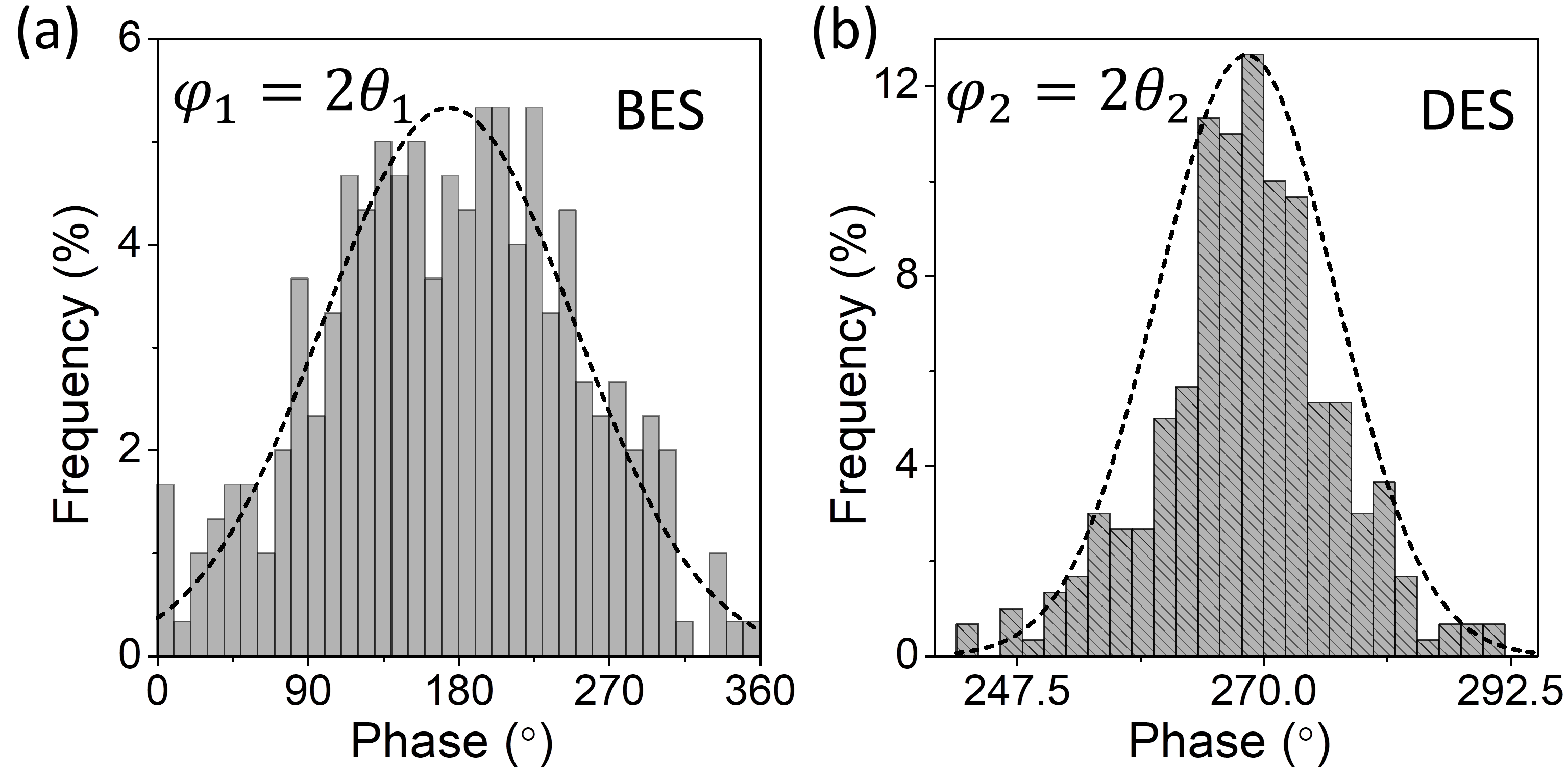}
  \end{center}
  \caption{Histogram of (a) $2\theta_1$ and (b) $2\theta_2$ angles using the notation of Eq.~\ref{Hexchrot}.}
  \label{fig:phases}
\end{figure}

Fig.~\ref{fig:phases} shows the histogram of exchange interaction phases $\theta_1$ and $\theta_2$ angles using the notation of Eq.~\ref{Hexchrot}. Should alloying be neglected (in the $C_{2v}$ case), $\phi_1=2\theta_1$ would be equal exactly 180$\degree$, and $\theta_2$ to 270$\degree$ revealing 90$\degree$ phase difference between dark and bright exciton. As seen in Fig.~\ref{fig:phases} when alloying is accounted for, this relation is held approximately, with strong bright exciton phase randomization.

\section{Two channels of dark excitons optical activity}
For DE states, Eq.~\ref{Eq:lifetime} for the $s$-shell case only, and in-plane ($x$, $y$) polarizations, (and skipping energy index for simplicity) takes the form
\begin{equation}
\begin{split}
I^{\mathrm{s-shell}}= \Big|
& C_{e\downarrow,h\Uparrow} M^{x,y}_{e\downarrow,h\Uparrow}+
C_{e\uparrow,h\Downarrow} M^{x,y}_{e\uparrow,h\Downarrow}+\\
& C_{e\uparrow,h\Uparrow} M^{x,y}_{e\uparrow,h\Uparrow}+
C_{e\downarrow,h\Downarrow} M^{x,y}_{e\downarrow,h\Downarrow}
\Big|^2,
\end{split}
\label{Eq:intensity_s}
\end{equation}
where $C$'s are expansion coefficients from diagonalization of the $4\times4$ CI Hamiltonian expressed in the $s$-shell basis.

When exchange mixing terms are neglected ($\mathbf{H}_{\mathrm{BD}}=0$) this formula is simplified with only with dark exciton optical activity possible if $M^{x,y}_{e\uparrow,h\Uparrow}\neq 0$, and
$M^{x,y}_{e\downarrow,h\Downarrow}\neq 0$, thus due to non-vanishing optical dipole moment corresponding to nominally dark configurations:
\begin{equation}
I^{\mathrm{s-shell}}_{\mathrm{dipole}}= \Big|
C_{e\uparrow,h\Uparrow} M^{x,y}_{e\uparrow,h\Uparrow}+
C_{e\downarrow,h\Downarrow} M^{x,y}_{e\downarrow,h\Downarrow}
\Big|^2.
\label{Eq:intensity_s_dipole}
\end{equation}
On the other hand, when exchange mixing terms are accounted for, while in-plane optical dipole moments are neglected, i.e.: $M^{x,y}_{e\uparrow,h\Uparrow}=M^{x,y}_{e\downarrow,h\Downarrow}=0$, dark exciton optical activity is possible via an admixture of oscillator strengths from bright configurations ($C_{e\downarrow,h\Uparrow}\neq 0$ or $C_{e\uparrow,h\Downarrow}\neq 0$)
\begin{equation}
I^{\mathrm{s-shell}}_{\mathrm{exch}}= \Big|
C_{e\downarrow,h\Uparrow} M^{x,y}_{e\downarrow,h\Uparrow}+
C_{e\uparrow,h\Downarrow} M^{x,y}_{e\uparrow,h\Downarrow}
\Big|^2
\label{Eq:intensity_s_mixing}.
\end{equation}
Both effects apparently play comparable roles, as discussed in the main text, and must be accounted for on equal footing by using Eq.~\ref{Eq:intensity_s} for the $s$-shell case only, or by using Eq.~\ref{Eq:intensity} when $s$, $p$, and $d$ shells are accounted for.

\bibliography{main}
\end{document}